\documentclass[fleqn,twoside]{article} 
\usepackage{graphicx}
\def\beq{\begin{equation}}
\def\eeq{\end{equation}}
\def\beqa{\begin{eqnarray}}
\def\eeqa{\end{eqnarray}}

\begin{document}

\begin{flushright}
JLAB-THY-98-07\\
NCKU-HEP-98-04\\
WM-98-103\\
\end{flushright}
\bigskip

\pagestyle{myheadings}

\centerline{\Large EXCLUSIVE PROCESSES AT INTERMEDIATE ENERGY, } 
\medskip
\centerline{ \Large QUARK-HADRON DUALITY}
\medskip
\centerline{\Large AND THE TRANSITION TO PERTURBATIVE QCD}

\bigskip

\centerline{\large {Claudio Corian\`{o}}}

\medskip
\centerline{Theory Group, Jefferson Lab, Newport News, VA 23606, USA}
\centerline{E-mail: coriano@jlab.org}
\medskip
\centerline{\large {Hsiang-nan Li}}

\medskip
\centerline{Department of Physics, National Cheng-Kung University}
\centerline{Tainan, Taiwan, Republic of China}
\centerline{E-mail: hnli@mail.ncku.edu.tw}

\medskip
\centerline{\large{ Cetin Savkl\i}}
\centerline{Department of Physics, College of William and Mary,}
\centerline{Williamsburg, VA 23187, USA}
\centerline{E-mail: csavkli@physics.wm.edu}

\bigskip
\noindent

{\sc key words}:\quad Exclusive Processes, Factorization Theorems,
Sudakov Resummation, QCD Sum Rules, Compton Scattering.

\textheight  612pt \textwidth  432pt
\headheight  12pt \headsep  20pt

\bigskip
\hrule
\bigskip
\begin{abstract}

Experiments at CEBAF will scan the intermediate-energy region of the QCD
dynamics for the nucleon form factors and for Compton Scattering. 
These experiments will definitely clarify the role of resummed
perturbation theory and of quark-hadron duality (QCD sum rules) in this
regime. With this perspective in mind, we review the factorization theorem
of perturbative QCD for exclusive processes at intermediate energy scales,
which embodies the transverse degrees of freedom of a parton and the Sudakov
resummation of the corresponding large logarithms. We concentrate on the
pion and proton electromagnetic form factors and on pion Compton scattering.
New ingredients, such as the evolution of the pion wave function and the
complete two-loop expression of the Sudakov factor, are included.
The sensitivity of our predictions to the infrared cutoff
for the Sudakov evolution is discussed. We also elaborate on QCD sum rule
methods for Compton Scattering, which provide an alternative description of
this process. We show that, by comparing the local duality analysis
to resummed perturbation theory, it is possible to describe the 
transition of exclusive processes to perturbative QCD. 
 
\end{abstract}

\newpage
\tableofcontents

\section{Introduction}

With the advent of CEBAF, the Continuous Electron Beam Accelerator Facility
at Jefferson LAB, studies of the intermediate-energy QCD and analyses
of various phenomena in hadronic physics, have now become a reality. 
These studies are performed experimentally with a very high
beam luminosity  through scattering of electrons on fixed targets.

Among these, just to mention few of them, are the investigations of the
hadronic resonances which are not classified by the ``traditional'' quark
model, such as the $J^{PC}$ exotic states that are not accessible to any
perturbative QCD (PQCD) prediction, and the study of the behaviour of
elastic scattering at the low end of the intermediate-energy region. By
``low-end'' we refer here to the few-GeV region ($Q^2\approx 1-3$ GeV$^2$),
where PQCD does not apply, and in general, 
one has to resort to various models to describe the data. 

We should also mention that there is the possibility that
an upgrade of CEBAF will bring $Q^2$ up into the bulk of the
intermediate-energy region, where the degrees of freedom of quarks and
gluons become more accessible to a perturbative treatment, using the
parton model. 

Our aim, in this paper, is to overview, from our perspective, the
phenomenology of elastic scattering off pions and nucleons, and to summarize
the open problems, which are left for future investigations.  We shall focus
the attention on the behaviour of form factors and on Compton scattering. 
An experimental overview of these studies can be found in Ref.~\cite{HW}.

This article is entirely based on the results obtained in the last few years 
in \cite{CRS,LS,CC1,CC2,CL,coli,L} with the addition 
of new results, on which we elaborate in great detail.  

We remark that excellent reviews discussing
the physics of form factors at intermediate energy have appeared recently
\cite{GSPS}, to which we refer for an overview of the subject.  

It is important to remark that the analysis of 
form factors at even lower energies requires the solution of bound state
equations in the Dyson-Schwinger or Bethe-Salpeter formalism. Recent
progress in the numerical study of form factors from the viewpoint
of the Dyson-Schwinger formalism has been reported in \cite{CT}. 

In the first part of this article, we elaborate in great detail on the
PQCD description of the pion and proton form factors, and extend the
discussion to the case of Compton scattering at intermediate energy.
New ingredients, such as the evolution of the pion wave function and the
complete two-loop expression of the Sudakov factor, are added. We show
that with the inclusion of these ingredients, the experimental data of the form
factors at intermediate energy can be explained.

Afterwards, we focus our discussion mainly on the connection between resummed
perturbation theory and QCD sum rules in the case of Compton scattering.

\subsection{Comparing Two Approaches: Sum Rules versus PQCD}
The applicability of PQCD in any process requires a hard scattering 
contribution. In the case of the pion and proton wave functions, 
for instance, the formation of the final bound state is mediated by the 
hard scattering and we need large energy and momentum 
transfer for this to happen. This is the so called ``asymptotic limit'', 
in which there is no direct overlap between the initial and the final state.
However, 
there is widespread agreement that at current energy the description is sub-asymptotic and a pure perturbative description is not the correct thing to do. 
If this is the case, then the correct physical mechanism underlying the 
dynamics of elastic scattering in the few Gev region is the direct overlap of wave functions, 
which has to be described in some other way. One possibility to describe such 
``soft'' behaviour is by a dispersion relation 
together with an Operator Product Expansion (quark-hadron duality). 
The combination of these two tools is known under the name of QCD sum rules.

A preliminary local duality description of real and virtual Compton
scattering was developed in \cite{CRS} in the analysis of fixed angle
Compton scattering, where a sum rule for the pion was derived. The study of
the power corrections and of the analiticity properties of the dispersion
relations for this process was presented in \cite{CC1,CC2}. Subsequently, an
analysis of Compton scattering by using resummed perturbation theory and QCD
sum rules was proposed by two of us as a way to get information on the
behaviour of the process at intermediate energy \cite{CL}. It was observed
that the two approaches generate different predictions for pion Compton
scattering at the high end and at the low end of the intermediate-energy 
region, with an interesting interplay between them. It is then possible to determine
the transition region to PQCD in pion Compton scattering by comparing these
predictions. Though the analysis was limited to the pion case, the extension
of this study to the proton case is straightforward. Work in this direction
is now in progress.

The results of \cite{CL} indicate that sum rule predictions for the
cross section of pion Compton scattering dominate over the resummed
perturbative predictions at lower momentum transfers. At larger momentum
transfers, instead, the perturbative predictions dominate over the sum rule
ones. The analysis in \cite{CL} also showed that the PQCD predictions
are very sensitive to variations in the momentum transfer $t$, while the
sum rule predictions are less sensitive to it.

If this interplay is indeed confirmed -as we expect- in more complex
processes, such as virtual Compton scattering off protons (VCS), we shall
have a lesson to learn: the transition to PQCD in VCS or in the real case
(RCS) may take place at a lower scale.
These conclusions, unfortunately, are affected by the usual model dependence
from the nonperturbative parametrizations involved in both resummed
perturbation theory and in QCD sum rules. However, the implications are
still important and can help us understand the experimental results.

On the other hand, the applicability of PQCD to exclusive processes has
been a matter of controversy for almost two decades \cite{IL1,HS}. Although
there is a general agreement that PQCD can successfully predict the
behaviour of exclusive reactions as the momentum transfer goes to infinity,
it remains unclear whether experimentally accessible energy scales
(at CEBAF) are large enough to justify these predictions. In this work we
briefly review the progress on this subject starting from the 80's, and
concentrate on the resummed PQCD formalism developed in \cite{BS,LS} for
elastic scattering at intermediate energy. This formalism modifies the
conventional factorization formulas for elastic hadron scattering at large
angle, which is the origin of the quark counting rules, by
including the Sudakov effects. These effects are important both in
exclusive and inclusive processes in special kinematic configurations in
which a small transverse momentum generates large logarithms that need to
be resummed to all orders. It was found that the effect of the 
resummation results 
in a suppression on the non-perturbative contributions, and improves the
applicability of PQCD at the high end of the intermediate-energy
region. It certainly remains legitimate to ask if the Sudakov effects are
really important at, say, $Q^2\approx 5$ GeV$^2$, or negligible as claimed
in \cite{MR} recently.

\subsection{QCD Sum Rules and Compton Scattering}

Among the semi-phenomenological methods developed for the study  of the
resonant region of QCD, sum rules are closest to a fundamental description
of the strong interactions. They have been formulated in a form 
which is very similar to PQCD. In the sum rule approach a ``master
equation" relates the timelike region of external invariants in 
correlation functions, the region where the resonances are located, to the
Euclidean region, where information from the Operator Product Expansion
(OPE) is parameterized in terms of local vacuum condensates (power
corrections). In practical applications only the phenomenological input
from the lowest dimensional (quark and gluon) condensates is required. 
The information on the resonance region, or timelike region, instead, is
modeled into a spectral density of states, which is described by the usual
ansatz ``lowest pole plus continuum''. 

A superconvergence condition requires the removal of the continuum
contribution  beyond a given ``duality interval". In fact, equating the
model spectral density to the OPE series leads to the cancellation of
the continuum contributions, {\it i.e.}, the contributions from both the
left-hand side and the right-hand side of the sum rule \cite{coli}.
The consequence of this procedure is a sum rule, which
relates properties of the lowest resonance to an OPE series that includes
both perturbative and nonperturbative terms. Radiative corrections to
spectral functions can also be calculated \cite{CC2,CJIR}. 

This way to obtain information about the resonances, which works very well
for both 2- and 3-point processes, has largely contributed to our
understanding of QCD in the intermediate-energy region \cite{SVZ,NR1,JS}.
In the extension of the method to 4-point processes, various new
difficulties have been encountered, such as more complex dispersion
relations, and the issue of the validity of the OPE. As discussed in
\cite{CRS}, the latter condition sets a constraint on the allowed range of
the variables $s$ and $t$, $s$ being the center-of-mass energy, and demands
to limit the analysis to scatterings at fixed-angle (or $-t/s$ equivalently). 

Once these kinematic constraints are satisfied, {\em new} information
available from sum rules for 4-point processes, compared to 3-point ones,
comes from the more general $s$ and $t$ dependences of the residui at
double poles of the lowest resonance, which are related to a perturbative
spectral function plus power corrections. Therefore, in the case of 4-point
processes a true angular dependence of the sum rules comes in to play, and
a parallel study of Compton scattering using this method \cite{CRS} and the
resummed PQCD factorization formalism is possible. This aspect has been
preliminarily discussed in \cite{CL}.

We review the above two methods that were used in the investigation of
elastic scatterings, concentrating on the pion and proton electromagnetic
form factors, and pion Compton scattering. The standard PQCD theory for
exclusive processes is summarized in Sect. 2. The Sudakov resummation,
which will be employed throughout the paper, is reviewed in Sect. 3, by
taking the pion form factor as an example. The technique can be trivially
extended to the proton case. The pion and proton form factors and pion
Compton scattering are evaluated in Sects. 4, 5, and 7, respectively. We
analyze pion Compton scattering using QCD sum rules in Sect. 6. Section 8 
contains our conclusion. The Appendices contain some details of the calculations.

\section{Perturbative QCD in Exclusive Processes}

\subsection{The Efremov-Radyushkin-Brodsky-Lepage Theory}

The PQCD theory for exclusive processes was first proposed by Brodsky and
Lepage \cite{LB1} and by Efremov and Radyushkin \cite{ER}. 
These authors argued that an exclusive process, such as the pion
electromagnetic form factor, can be factorized into two types of
subprocesses: wave functions which carry the nonperturbative information of
the initial and final state pions, and a hard amplitude which
describes the scattering of a valence quark of the pion by the energetic
photon. The former is not calculable in perturbation theory, and needs to
be parametrized by a model, to be derived by nonperturbative methods, such
as QCD sum rules and lattice gauge theory, or to be determined by
experimental data. The latter, characterized by a large momentum flow, is
calculable in perturbation theory. Combining these two types of
subprocesses, predictions for the pion form factor at experimentally
accessible energy scales can be made.

According to this standard picture, the factorization formula for the pion
form factor $F_\pi(Q^2)$, graphically represented by Fig.~1(a), is written
as
\begin{equation}
F_\pi(Q^2)=\int_0^1 d x_1d x_2\phi(x_2,\mu)H(x_1,x_2,Q,\mu)\phi(x_1,\mu)\;,
\label{pi}
\end{equation}
$Q^2=-2P_1\cdot P_2$ being the momentum transfer from the photon, and $P_1$
($P_2$) the mometum of the incoming (outgoing) pion. $\mu$ is a
renormalization and factorization scale, below which the QCD dynamics is
absorbed into the pion wave function $\phi$, and above
which the dynamics is regarded as perturbative and absorbed into the hard
amplitude $H$. The pion wave function $\phi(x,\mu)$ gives the probability
of a valence quark carrying a fractional momentum $xP$ in the parton model
at the energy scale $\mu$. Usually, the $\mu$ dependence of $\phi$ is
neglected. $H$ is obtained by computing the quark-photon scattering diagrams
as shown in Fig.~1(b). All the lines in $\phi$ are thought of as near or on
the mass shell, while those in $H$ are far off-shell by the scale $Q$.

To make predictions for $F_\pi$, one substitutes the lowest-order expression
of $H$ obtained from Fig.~1(b),
\begin{equation}
H(x_1,x_2,Q,\mu)=\frac{16\pi {\cal C}_F \alpha_s(\mu)}{x_1x_2Q^2}\;,
\label{pih}
\end{equation}
$\alpha_s(\mu)$ being the running coupling constant, and the asymptotic
pion wave function \cite{LB2},
\begin{equation}
\phi^{AS}(x)=\frac{3f_{\pi}}{\sqrt{2N_{c}}}x(1-x)\;,
\label{as}
\end{equation}
into Eq.~(\ref{pi}), where ${\cal C}_F=4/3$ is a color factor, $N_{c}=3$
the number of colors, and $f_{\pi}=0.133$ GeV the pion decay constant.
Because of the behaviour of $\phi^{AS}$, the main contributions to
Eq.~(\ref{pi}) come from the regions with intermediate $x_1$ and $x_2$.
Higher-order corrections to $H$ then produce the logarithms of the type
$\ln(Q^2/\mu^2)$, which may be so large as to spoil the perturbative
expansion. To eliminate these logarithms, a natural choice of $\mu^2$ is
$\mu^2=Q^2$. It is easy to obtain $Q^2F_\pi\sim 0.12$ GeV$^2$ for
$Q^2\sim 4$-10 GeV$^2$, which is only 1/3 of the data 0.35 GeV$^2$
\cite{d1,d2,d3,CJB}. This contradiction implies
that the pion form factor has not yet become completely asymptotic at
experimentally accessible energy scales.

Hence, one may resort to a preasymptotic model, the Chernyak and Zhitnitsky
(CZ) wave function derived from QCD sum rules \cite{CZ1},
\begin{equation}
\phi^{CZ}(x)=\frac{15f_{\pi}}{\sqrt{2N_{c}}}x(1-x)(1-2x)^{2}\;,
\label{cz}
\end{equation}
which possesses maxima at the end points $x\to 0$ and $x\to 1$. It was found
that the CZ wave function enhances the predictions for $Q^2F_\pi$ to
0.30 GeV$^2$, and improves the match with the data.

In spite of the success of the standard PQCD theory, the
perturbative expressions for the pion form factor were brought under
searching criticism by Isgur and Llewellyn Smith \cite{IL1} and by
Radyushkin \cite{BR}. Since $\phi^{CZ}$ emphasizes the contributions from
small $x_1$ and $x_2$, the end-point regions are important. They argued
that higher-order corrections to $H$ in fact produce logarithms like
$\ln(x_1x_2Q^2/\mu^2)$ and thus the natural choice of $\mu^2$ is
$\mu^2=x_1x_2Q^2$. For intermediate $x_1$ and $x_2$, this choice is
consistent with $\mu^2=Q^2$ employed above. For small $x_1$ and $x_2$, it
results in $\alpha_s(x_1x_2Q^2)>1$, indicating that the end-point regions
are nonperturbative regions. A careful analysis reveals that the
enhancement of the predictions for $Q^2F_\pi$ by $\phi^{CZ}$ is due to
the amplification of the nonperturbatve end-point contributions. As a
consequence, the perturbative calculation loses its
self-consistency as a weak-coupling expansion.

A cutoff to avoid the divergences of $\alpha_s$ was proposed \cite{JSL}:
\begin{equation}
\alpha_{s}(\mu)=\frac{\pi}{\beta_0 \ln
[(\mu^{2}+4m_{g}^{2})/\Lambda^{2}]}\;,
\label{2}
\end{equation}
with $\Lambda\equiv \Lambda_{\rm QCD}$ the QCD scale parameter and
$\beta_0=(33-2n_f)/12$, $n_f=3$ being the number of quark flavors. $m_{g}$
is interpreted as a dynamical gluon mass, which is acquired from the low
momentum region of those radiative corrections that induce the running
of $\alpha_s$ \cite{JMC}. The leading-order PQCD predictions are then
stabilized at low momentum transfers because of the frozen $\alpha_s$.
In this interesting approach the value of $m_{g}$ has to be determined by
matching the predictions with the data. However, the results from
Eq.~(\ref{pi}) are slightly sensitive to the variation of $m_{g}$ around
the best choice $m_{g}^{2}\approx 0.3$ GeV$^{2}$ \cite{JSL}, and thus the
theory with this cutoff becomes less predictive. An alternative way to freeze 
the coupling constant at low momentum transfer is the BLM method, in which the 
renormalization scale $\mu$ is fixed according to some criteria. 
We refer to \cite{BBB} for more details on the application of this method 
to exclusive processes. 

\subsection{Resummed PQCD Formalism}

The above discussions hint that the end-point regions should be treated in a
different way. In general, a valence quark in the pion can carry a small
amount of transverse momenta $k_T$. For intermediate $x_1$ and $x_2$,
$k_T$ that flow from the wave functions through the hard scattering
are neglected as power-suppressed corrections, as shown by the hard gluon
propagator
\begin{equation}
\frac{1}{x_1x_2Q^2+k_T^2}\approx
\frac{1}{x_1x_2Q^2}\left(1-\frac{k_T^2}{x_1x_2Q^2}\right)\;.
\end{equation}
The first term is exactly the asymptotic hard scattering in Eq.~(\ref{pih}),
and the second is suppressed by powers of $Q^{2}$ at fixed $x$. If the
end-point regions are not important, one may drop the second term. However,
the end-point difficulties indicate that the higher-power effects are
crucial, and should be kept at the outset in the derivation of the new
perturbative expression for the pion form factor.

The introduction of the $k_T$ dependence leads to three observations
immediately. First, the pion form factor becomes a two-scale ($Q$ and $k_T$)
problem, and the resummation technique may be required to organize the large
logarithms $\ln(Q/k_T)$ from radiative corrections contained in the wave
function. Second, we must analyze the process in the Fourier transform space
of $k_T$, denoted by $b$, which is regarded as the transverse
separation between the valence quarks of the pion. Third, the quantity
$1/b$, now considered as one of the characteristic scales of the hard
scattering amplitude, should be substituted for the argument of $\alpha_s$
if $1/b^2 > x_1x_2Q^2$. The nonperturbative regions are then characterized
by the regions of large $b$ and small $x$.

We find that the resummation of the large logarithms behaves as
\begin{equation}
\exp[-{\rm const.}\times\ln Q\ln(\ln Q/\ln b)]\;,
\label{sqb}
\end{equation}
which suppresses the elastic
scattering at large spatial separation. This property, called Sudakov
suppression \cite{BS,CS,JCC}, makes the nonperturbative contributions
from large $b$, no matter what $x$ is, less important, without introducing
any phenomenological parameters such as a gluon mass $m_g$. The 
relevance of the Sudakov
effects was pointed out by Lepage and Brodsky in \cite{LB1}. The
underlying physical principle is that the elastic scattering of an isolated
colored parton, such as a quark, is suppressed at high energy by radiative
corrections. We shall show that the new perturbative expression reduces to
the standard one in Eq.~(\ref{pi}) as $Q\rightarrow \infty$, but, at lower
momentum transfers, it takes into account the infinite summation of those 
higher-order effects whose $b$ dependence makes the perturbative theory
more self-consistent -at least at the high end of the experimentally accessible
energies-. This self-consistency means that numerical results come
predominantly from the momentum regions in which the effective coupling is
relatively small. The above argument can be quantified by studying the
dependence of the contribution to the form factor on a cutoff in $b$,
instead of in $x$ \cite{IL1,R}, from which the importance of the
perturbative region is examined. It is observed that the inclusion
of Sudakov corrections produces numerical effects very similar to the
approaches with a frozen $\alpha_s$ at low energies, and in a good
agreement with the data. A brief review of the standard asymptotic
expression for the form factors is referred to \cite{ER,LB2,CZ1,JSL,FJ,CG}.
The perturbative calculations should be thought of as complementing the QCD
sum rule approach \cite{NR1}, which is one of the topics of this paper.

\section{Sudakov Resummation}

As we have discussed above, the applicability of PQCD to exclusive processes
at intermediate energy can be improved by including the transverse
momentum dependence and the Sudakov suppression into the 
factorization formulas. 
In this section we review the resummation technique, taking the pion form
factor as an example. We investigate radiative corrections to the basic
diagrams for the form factor, and explain how they are factorized.
For details on the Sudakov resummation, we refer to \cite{CS}.

\subsection{Factorization in $b$ Space}

To construct the factorization of QCD processes we first isolate the
leading momentum regions of those radiative corrections from which important
contributions to the loop integrals arise. There are two types of important
contributions: collinear, when the loop momentum is parallel to the
pion momentum, and soft, when the loop momentum is much smaller than 
the momentum transfer $Q^2$. A small amount of transverse momenta $k_T$
is associated with the valence partons that enter hard scattering 
which is taken as an infrared cutoff.
Each type of the important (soft or collinear) contributions gives large
single logarithms, and their overlap generates double (leading) logarithms,
which are a characteristic feature of gauge theories. These large
logarithms, appearing in a product with $\alpha_s$, must be organized
in order not to spoil the perturbative expansion. Single logarithms can be
summed to all orders using renormalization group (RG) methods, while double
logarithms must be treated by the resummation technique \cite{CS}.

The diagrams shown in Fig.~2 represent $O(\alpha_s)$ corrections to the
basic factorization of the pion form factor, containing the large logarithms
mentioned above. In the axial (physical) gauge $n\cdot A=0$, $n$ being a
gauge vector and $A$ the gauge field, the two-particle reducible diagrams
in the channels of the external pions, like Figs.~2(a) and 2(b), have
double logarithms from the overlap of collinear and soft divergences, while
the two-particle irreducible corrections, like Figs.~2(c) and 2(d), contain
only single soft logarithms. This distinction is consistent with the
physical picture of a hard scattering process: two partons moving in the same
direction can interact with each other through collinear or soft gluons,
while those moving apart from each other can interact only through soft
gluons. The Sudakov corrections come entirely from diagrams that are
two-particle reducible. Below we shall concentrate on reducible corrections,
and demonstrate how they are summed into a Sudakov factor.

A careful analysis shows that soft divergences cancel between Figs.~2(a) and
2(b), as well as between 2(c) and 2(d), in the asymptotic region with
the transverse separation $b\to 0$. Therefore, reducible corrections are
dominated by collinear divergences, and can be absorbed into the pion wave
funtion, which involves a similar dynamics. Irreducible corrections,
due to the cancellation of soft divergences, are then absorbed into the
hard scattering amplitude $H$. Hence, the factorization picture holds at
least asymptotically after the radiative corrections are included. The
cancellation of soft divergences is closely related to the universality of
wave functions. For a large $b$, double logarithms are present and the
resummation technique must be implemented.

When the transverse degrees of freedom of a parton are included, the
factorization of the reducible corrections into the pion wave function must
be performed in $b$ space. Fig.~2(a), where the loop momentum $l$ does
not flow through the pion wave function, is computed directly. For
Fig.~2(b), where $l$ may be routed through the wave function $\psi$, the
associated integrand contains a factor
$\psi(x_1+l^+/P_1^+,|{\bf k}_{1T}+{\bf l}_T|)$.
Here we have assumed that $l^-$ is routed through the outgoing pion wave
function. Since the behavior of $\psi$ is unknown, the loop integral can
not be done directly. We employ the approximation $\psi(x_1+l^+/P_1^+,
|{\bf k}_{1T}+{\bf l}_T|)\approx \psi(x_1,|{\bf k}_{1T}+{\bf l}_T|)$, since
$P_1^+$ is large and the $l^+$ dependence is negligible. However, the
transverse momentum $k_{1T}$ is not large, and thus the $l_T$ dependence
remains. This difficulty can be overcome by introducing a Fourier transform,
\begin{equation}
\psi(x_1,|{\bf k}_{1T}+{\bf l}_T|)=\int\frac{d^2{\bf b}}{(2\pi)^2}
\exp[-i({\bf k}_{1T}+{\bf l}_T)\cdot {\bf b}]{\cal P}(x_1,b)\;,
\end{equation}
with $\cal P$ the Fourier transformed wave funciton.
The factor $\exp(-i{\bf l}_T\cdot {\bf b})$ is absorbed into the loop
integrand, which then becomes computable, and the factor 
$\exp(-i{\bf k}_{1T}\cdot {\bf b})$ Fourier transforms the hard part $H$ to $b$ space.

Using the above reasoning, the factorization formula for the pion form
factor is written as \cite{LS}
\begin{eqnarray}
F_{\pi}(Q^{2})&=&\int_0^1 d x_{1}d x_{2}\int
\frac{d^2 {\bf b}}{(2\pi)^2}
{\cal P}(x_{2},b,P_{2},\mu)
\nonumber \\
& &\times {\tilde H}(x_1,x_2,b,Q,\mu)
{\cal P}(x_{1},b,P_{1},\mu)\; ,
\label{pi1}
\end{eqnarray}
where the expression of the Fourier transformed hard amplitude ${\tilde H}$
will be given in next section. Equation (\ref{pi1}) depends only on a
single parameter $b$, because the virtual quark line involved in $H$ is
thought of as being far from mass shell, and shrunk to a point \cite{LS}.
The wave function ${\cal P}$ is defined in terms of explicit matrix
elements, which complement the diagrammatic descriptions presented in
Fig.~1(a) \cite{BS},
\begin{eqnarray}
{\cal P}(x,b,P,\mu)
&=&\int d^2 {\bf k}_T e^{i{\bf k}_T\cdot {\bf b}}\psi (x,k_T,P,\mu)
\nonumber \\
&=& \int {d y^- \over 2\pi}
e^{ixP^+y^-} \langle 0|T({\bar q}(0)\gamma^+ q(y^-,0^+,{\bf b}))|\pi(P)
\rangle\; .
\label{def}
\end{eqnarray}
${\cal P}$ includes all leading logarithmic
enhancements at large $b$, which will be resummed below.

Equation (\ref{pi1}) is an intermediate step in deriving Eq.~(\ref{pi})
\cite{LB2}. If one sets $b$ to zero in ${\cal P}$, and integrates over
$b$ in ${\tilde H}$, the former reduces to the latter. For fixed
$x_i\not= 0$, the two expressions are
equivalent up to corrections that fall off as a power of $Q$. However,
Eq.~(\ref{pi1}) retains much more higher-power information at the limit
$x_i\rightarrow 0$ for fixed $Q$, which is potentially important.

\subsection{Technique}

The basic idea of the resummation technique is as follows. If the double
logarithms appear in an exponential form in Eq.~(\ref{sqb}), the task will
be simplified by studying the derivative
$d{\cal P}/d\ln Q=C{\cal P}$. The coefficient function $C$ contains only
large single logarithms, and can be treated by RG methods. Therefore,
working with $C$ one simplifies the double-logarithm problem into a
single-logarithm problem.

The two invariants appearing in ${\cal P}$ are $P\cdot n$ and $n^2$. Due to
the structure of the gluon propagator in the axial gauge,
\begin{equation}
N^{\mu\nu}(l)=\frac{-i}{l^2}\left(g^{\mu\nu}-\frac{n^{\mu}l^{\nu}+
l^{\mu}n^{\nu}}{n\cdot l}+n^2\frac{l^{\mu}l^{\nu}}{(n\cdot l)^2}\right)\;,
\label{gpr}
\end{equation}
${\cal P}$ depends only on a single large scale $(P\cdot n)^2/n^2$.
We choose the Breit frame such that
$P_1^+=P_2^-=Q/\sqrt{2}$ and all other components of $P$'s vanish.
It is then easy to show that the differential operator $d/d\ln Q$ can
be replaced by $d/d n$,
\begin{equation}
\frac{d}{d \ln Q}{\cal P}=-\frac{n^2}{P\cdot n}P^{\alpha}
\frac{d}{d n^{\alpha}}{\cal P}\;.
\label{qn}
\end{equation}
The motivation for this replacement is that the momentum $P$ flows through
both quark and gluon lines, but $n$ appears only in gluon lines. The
analysis then becomes simpler by studying the $n$, instead of $P$,
dependence.

Applying $d/d n_{\alpha}$ to the gluon propagator, we obtain
\begin{equation}
\frac{d}{d n_{\alpha}}N^{\mu\nu}=-\frac{1}{l\cdot n}
(N^{\mu\alpha}l^{\nu}+N^{\nu\alpha}l^{\mu})\;.
\label{dn}
\end{equation}
The momentum $l^\mu$ ($l^\nu$) is contracted with the vertex  the
``differentiated'' gluon attaches, which is then replaced by a new vertex,
\begin{eqnarray}
gT^a\frac{n^2}{P\cdot n l\cdot n}P_{\alpha}\;,
\end{eqnarray}
$T^a$ being a color matrix. This new vertex can be easily read off from the
combination of Eqs.~(\ref{qn}) and (\ref{dn}). After adding
together all the diagrams with ``differentiated'' gluons and using
the Ward identity, we arrive at the differential equation of ${\cal P}$ 
shown in Fig.~3(a), in which the new vertex is represented by a square.
An important feature of the new vertex is that the gluon momentum $l$
does not lead to collinear divergences because of the nonvanishing $n^2$.
The leading regions of $l$ are then soft and ultraviolet, in which Fig.~3(a)
can be factorized according to Fig.~3(b) at lowest order. The part on the
left-hand side of the dashed line is exactly ${\cal P}$, and that on the
right-hand side is assigned to the coefficient function $C$.

Hence, we introduce a function ${\cal K}$ to absorb the soft divergences
from the first two diagrams in Fig.~3(b), and a function ${\cal G}$ to
absorb the ultraviolet divergences from the other two diagrams. The
soft subtraction employed in ${\cal G}$ ensures that the involved
momentum flow is hard. Generalizing the two functions to all orders,
we derive the differential equation of ${\cal P}$,
\begin{eqnarray}
\frac{d}{d \ln Q}{\cal P}(x,b,P,\mu)&=&\left[\,2{\cal K}
(b\mu)+{\cal G}(x\nu/\mu)+{\cal G}((1-x)\nu/\mu)\right]
\nonumber \\
& &\times{\cal P}(x,b,P,\mu)\; .
\label{qp}
\end{eqnarray}
${\cal K}$ and ${\cal G}$ have been calculated to one loop, and their
single logarithms have been organized to give the evolutions in $b$ and
$Q$, respectively \cite{BS}. They have ultraviolet poles individually,
but their sum ${\cal K}+{\cal G}$ is finite.

Substituting the expressions for ${\cal K}$ and ${\cal G}$ into
Eq.~(\ref{qp}), we obtain the solution
\begin{equation}
{\cal P}(x,b,P,\mu)=\exp\left[-s(xQ,1/b)-s((1-x)Q,1/b)\right]
{\bar{\cal P}}(x,b,\mu)\; .
\label{sp}
\end{equation}
The exponent $s(\xi Q,1/b)$ for $\xi=x$ and $1-x$ is written as \cite{BS}
\begin{equation}
s(\xi Q,1/b)=\int_{1/b}^{\xi Q/\sqrt{2}}\frac{d p}{p}
\left[\ln\left(\frac{\xi Q}{\sqrt{2} p}\right)
A(\alpha_s(p))+B(\alpha_s(p))\right]\;,
\label{fsl}
\end{equation}
where the anomalous dimensions $A$ to two loops and $B$ to one loop are
given by
\begin{eqnarray}
A&=&{\cal C}_F\frac{\alpha_s}{\pi}+\left[\frac{67}{9}-\frac{\pi^2}{3}
-\frac{10}{27}n_f+\frac{8}{3}\beta_0\ln\left(\frac{e^{\gamma_E}}{2}\right)
\right]\left(\frac{\alpha_s}{\pi}\right)^2\;,
\nonumber \\
B&=&\frac{2}{3}\frac{\alpha_s}{\pi}\ln\left(\frac{e^{2\gamma_E-1}}
{2}\right)\;,
\end{eqnarray}
with $\gamma_E$ the Euler constant. The two-loop expression of the running
coupling constant,
\begin{equation}
\frac{\alpha_s(\mu)}{\pi}=\frac{1}{\beta_0\ln(\mu^2/\Lambda^2)}-
\frac{\beta_1}{\beta_0^3}\frac{\ln\ln(\mu^2/\Lambda^2)}
{\ln^2(\mu^2/\Lambda^2)}\;,
\label{2la}
\end{equation}
will be substituted into Eq.~(\ref{fsl}), with the coefficients
\begin{eqnarray}
\beta_{0}=\frac{33-2n_{f}}{12}\;,\;\;\;\beta_{1}=\frac{153-19n_{f}}{24}\;.
\label{12}
\end{eqnarray}
To derive Eq.~(\ref{fsl}), a space-like gauge vector
$n\propto (1,-1,{\bf 0})$ has been chosen. We require the ordering of the
relevant scales to be $\xi Q > 1/b > \Lambda$ as indicated by the
integration of $p$ from $1/b$ to $\xi Q$ in Eq.~(\ref{fsl}). The QCD
dynamics below $1/b$ is regarded as being nonperturbative, and absorbed
into the initial condition ${\bar{\cal P}}(\xi,b,\mu)$.

\section{The Perturbative Pion Form Factor}

In this section we evaluate the pion form factor in the framework
developed above. Compared to the analysis in \cite{LS}, the more
complete Sudakov factor derived from the two-loop running coupling constant
$\alpha_s$ in Eq.~(\ref{2la}) will be employed and the evolution of the
pion wave function will be taken into account. The above resummed PQCD
formalism has been widely applied to various exclusive processes, such as
the time-like pion form factor \cite{GP}, proton-antiproton annihilation
\cite{H} and proton-proton Landshoff scatterings \cite{SS}.

\subsection{Evolution in $b$}

We continue the organization of the large logarithms contained in the
factorization formula for the pion form factor in Eq.~(\ref{pi1}).
The functions ${\bar{\cal P}}$ and $\tilde H$ still contain single
logarithms coming from ultraviolet divergences, which need to be summed
using the RG equations \cite{BS},
\begin{eqnarray}
& &{\cal D}{\bar{\cal P}}(x,b,\mu)=-2\gamma_q {\bar{\cal P}}(x,b,\mu)
\label{pee}\\
& &{\cal D}{\tilde H}(x_i,b,Q,\mu)=4\gamma_q{\tilde H}(x_i,b, Q,\mu)\; ,
\label{re}
\end{eqnarray}
with
\begin{eqnarray}
{\cal D}=\mu\frac{\partial}{\partial \mu}+\beta(g)\frac{\partial}
{\partial g}\;,
\end{eqnarray}
and $\gamma_q=-\alpha_s/\pi$ being the quark anomalous dimension in the
axial gauge. Solving Eq.~(\ref{pee}), the large-$b$ behavior of $\cal P$
is summarized as
\begin{eqnarray}
{\cal P}(x,b,P,\mu)&=&\exp\left[-s(xQ,1/b)-s((1-x)Q,1/b)
-2\int_{1/b}^{\mu} \frac{d\bar{\mu}}{\bar{\mu}}\gamma
_q(\alpha_s(\bar{\mu}))\right]
\nonumber \\
& &\times {\bar{\cal P}}(x,b,1/b)\;,
\label{pb}
\end{eqnarray}
where the arguments $b$ and $1/b$ of the initial condition ${\bar{\cal P}}$
denote the intrinsic and perturbative $b$ evolutions, respectively.
Below we shall ignore the intrinsic $b$ dependence, and assume
that ${\bar{\cal P}}=\phi(x,1/b)$.

The RG solution of Eq.~(\ref{re}) is given by
\begin{eqnarray}
{\tilde H}(x_i,b,Q,\mu)
=\exp\left[-4\,\int_{\mu}^{t}\frac{d\bar{\mu}}{\bar{\mu}}
\gamma_q(\alpha_s(\bar{\mu}))\right]{\tilde H}(x_i,b,Q,t)\;,
\label{13}
\end{eqnarray}
where $t$ is the largest mass scale involved in the hard scattering,
\begin{equation}
t=\max(\sqrt{x_{1}x_{2}}Q,1/b)\; .
\label{9}
\end{equation}
The scale $\sqrt{x_1x_2}Q$ is associated with the longitudinal momentum of
the hard gluon and $1/b$ with the transverse momentum. The anomalous
dimensions $\gamma_q$ allow us to take into account the strong couplings of
quarks when $t$ is small. With the large logarithms organized, the initial
condition ${\tilde H}(x_i,b,Q,t)$ can be taken to coincide with its
lowest-order expression, obtained by a Fourier transform of the hard
scattering amplitude $H$ in momentum space, as derived from Fig.~1(b),
\begin{eqnarray}
H &=&
\frac{16\pi\alpha_s{\cal C}_Fx_{1}Q^{2}}
{(x_{1}Q^{2}+{\bf k}_{1T}^{2}) (x_{1}x_{2}Q^{2}+
({\bf k}_{1T}-{\bf k}_{2T})^2)}
\label{hna} \\
&\approx &\frac{16\pi\alpha_s{\cal C}_F}
{x_{1}x_{2}Q^{2}+({\bf k}_{1T}-{\bf k}_{2T})^{2}}\;.
\label{ah}
\end{eqnarray}
We have neglected the transverse momentum in the numerator of the first
expression, and neglected, in addition, the transverse momentum
carried by the virtual quarks in the denominator of the second expression.
These quark propagators are linear rather than quadratic in $x$. Because
Eq.~(\ref{ah}) depends on the combination of the transverse momenta, the
factorization formula for the pion form factor involves only a
single-$b$ integral as shown in Eq.~(\ref{pi1}). Note that Eq.~(\ref{ah})
coincides with Eq.~(\ref{pih}) at ${\bf k}_{iT}=0$.

Inserting Eqs.~(\ref{pb}) and (\ref{13}) into Eq.~(\ref{pi1}), it becomes
\begin{eqnarray}
F_{\pi}(Q^2)&=& 16\pi{\cal C}_F\int_{0}^{1}d x_{1}d x_{2}
\phi(x_{1},1/b)\phi(x_{2},1/b)
\int_{0}^{\infty} bd b \alpha_{s}(t) K_{0}(\sqrt{x_{1}x_{2}}Qb)
\nonumber \\
& &\times \exp[-S(x_{1},x_{2},b,Q)]\; ,
\label{15}
\end{eqnarray}
with the complete Sudakov exponent
\begin{equation}
S(x_{1},x_{2},b,Q)=\sum_{i=1}^{2}\left[s(x_{i}Q,1/b)+s((1-x_{i})Q,1/b)
\right]+4\int_{1/b}^{t}\frac{d\bar{\mu}}{\bar{\mu}}
\gamma_q(\alpha_s(\bar{\mu}))\;.
\label{16}
\end{equation}
$K_{0}$ is the modified Bessel function of order zero, which is the
Fourier transform to $b$ space of the gluon propagator. Equation (\ref{15})
for the pion form factor, being a physical quantity, is explicitly
$\mu$-independent. This is the consequence of the RG analysis in
Eqs.~(\ref{pee}) and (\ref{re}), and the advantage of the resummed PQCD
formalism.

Another advantage of Eq.~(\ref{15}), compared to Eq.~(\ref{pi}), is the
extra $b$ dependence. For $x$ close to zero, {\it i.e.}, $x_1x_2Q^2<1/b^2$,
the scale of the radiative corrections is dominated by the transverse
distance bridged by the exchanged gluon. If this distance is small,
radiative corrections with the argument of $\alpha_s$ set to $t$ in the
hard scattering amplitude will be small, regardless of the values of $x$.
Of course, when $b$ is large and $x_1x_2Q^2$ is small, radiative
corrections are still large. However, the Sudakov factor $\exp(-S)$,
exhibiting a strong fall off at large $b$, vanishes for $b > 1/\Lambda$.
We expect that, since Eq.~(\ref{pi}) is the asymptotic limit of
Eq.~(\ref{15}), as $Q$ increases, the $b$ integral will become more and
more dominated by the small-$b$ contributions. If the Sudakov suppression
is strong enough, such that the main contributions to the factorization
formula come from the short-distance region, the perturbative expansion
will be rendered relatively self-consistent. Hence, we propose to analyze
Eq.~(\ref{15}), not by cutting off the $x$ integrals near their endpoints,
but by testing its sensitivity to the large-$b$ region.

The evolution of the wave function $\phi$ with $b$ is written as \cite{Ji},
\begin{equation}
\phi(x,1/b)=\frac{3f_\pi}{\sqrt{2N_c}} x(1-x)
\left[\sum_{n=0}^{\infty}a_nC^{3/2}_n(2x-1)
\left(\frac{\alpha_s(1/b)}{\alpha_s(\mu_0)}\right)^{\gamma_n}
\right]\;,
\label{phie1}
\end{equation}
with $\mu_0\approx 0.5$ GeV. The coefficient $a_n$ and the exponent
$\gamma_n$ are
\begin{eqnarray}
a_n&=&\frac{5(2n+3)}{(n+2)(n+1)}I_n\;,
\\
\gamma_n&=&\frac{4}{33-2n_f}\left[1+4\sum_{k=2}^{n+1}\frac{1}{k}-
\frac{2}{(n+1)(n+2)}\right]\;,
\end{eqnarray}
with $I_0=2/15$, $I_1=0$, $I_2=8/35$ and $I_n=0$ for $n\ge 3$. The relevant
Gegenbauer polynomials are
\begin{equation}
C^{3/2}_0(x)=1\;,\;\;\;\;
C^{3/2}_2(x)=\frac{15}{2}\left(x^2-\frac{1}{5}\right)\;.
\end{equation}
That is, we adopt the pion wave function
\begin{equation}
\phi(x,1/b)=\frac{3f_\pi}{\sqrt{2N_c}}x(1-x)
\left[1+(5(2x-1)^2-1)\left(\frac{\alpha_s(1/b)}
{\alpha_s(\mu_0)}\right)^{\frac{50}{81}}\right]\;.
\label{phie}
\end{equation}
It is easy to observe that Eq.~(\ref{phie}) approaches $\phi^{AS}$ in
Eq.~(\ref{as}) as $1/b \to \infty$, and $\phi^{CZ}$ in Eq.~(\ref{cz}) as
$1/b \to \mu_0$. This is the reason $\phi^{AS}$ is an asympotic wave
function, while $\phi^{CZ}$ is a preasympotic one.

We are also interested in the full expression of $H$ in Eq.~({\ref{hna}),
except for the approximate one in Eq.~(\ref{ah}). The corresponding
factorization formula involves a double-$b$ integral
\begin{eqnarray}
F_{\pi}(Q^2)&=&16\pi{\cal C}_F \int_{0}^{1}d x_{1}d x_{2}
\int_{0}^{\infty} b_{1}d b_{1}b_{2}d b_{2}\phi(x_{1},1/b_1)\phi(x_{2},1/b_2)
\nonumber \\
& &\times\alpha_{s}(t)K_{0}(\sqrt{x_{1}x_{2}}Qb_{1})
\exp\left[-S(x_{1},x_{2},b_{1},b_{2},Q)\right]
\nonumber \\
& &\times \left[\theta(b_{1}-b_{2})
K_{0}(\sqrt{x_{1}}Qb_{1})I_{0}(\sqrt{x_{1}}Qb_{2})\right.+
\nonumber \\
& &\left.\theta(b_{2}-b_{1})K_{0}(\sqrt{x_{1}}Qb_{2})I_{0}(\sqrt{x_{1}}Qb_{1})
\right]\; ,
\label{f5}
\end{eqnarray}
with the Sudakov exponent
\begin{eqnarray}
S&=&\sum_{i=1}^2\left[s(x_{i}Q,1/b_{i})
+s((1-x_{i})Q,1/b_{i})\right]
\nonumber \\
& &+2\int_{1/b_1}^{t}\frac{d\bar{\mu}}{\bar{\mu}}
\gamma_q(\alpha_s(\bar{\mu}))
+2\int_{1/b_2}^{t}\frac{d\bar{\mu}}{\bar{\mu}}
\gamma_q(\alpha_s(\bar{\mu}))\;,
\end{eqnarray}
and the hard scale
\begin{equation}
t={\rm max}\left[\sqrt{x_1x_2}Q, 1/b_1, 1/b_2\right]\; .
\end{equation}

\subsection{Numerical Results}

Results on the behaviour of the Sudakov suppression in the large $b$ region 
where presented in \cite{LS}. There is no suppression for small $b$, where the
two fermion lines are close to each other. In this region the higher-order
corrections should be absorbed into the hard amplitude, instead of being
adsorbed into the pion wave function. Hence, we set any factor
$\exp[-s(\xi Q,1/b)]$ to unity for $\xi Q/\sqrt{2} < 1/b$.
Similarly, we also set the complete Sudakov exponential $\exp(-S)$ in
Eq.~(\ref{15}) to unity, whenever it exceeds unity in the small-$b$ region.
As $b$ increases, $\exp(-S)$ decreases, reaching zero at $b=1/\Lambda$.
Suppression in the large $b$ region is stronger for larger $Q$.

We choose the QCD scale $\Lambda=0.15$ GeV for the numerical analysis
of the pion form factor below. It can be shown that the results will
differ only by 10\%, if $\Lambda=0.1$ GeV or $\Lambda=0.2$ GeV is adopted.

In order to see how the contributions to Eq. (\ref{15}) are distributed in
the $b$ space under Sudakov suppression, the integration is done with $b$
cut off at $b_{c}$. Typical numerical results are displayed in Fig.~4
for the use of $\phi$ in Eq.~(\ref{phie}). The curves, showing the
dependence of $Q^{2}F_{\pi}$ on $b_{c}$, rise from zero at $b_c=0$, and
reach their full height at $b_{c}=1/\Lambda$, beyond which any remaining
contributions are regarded as being truly nonperturbative. To be quantitive,
we determine a cutoff up to which half of the whole contribution has
been accumulated. For $Q^2=4$ and 9 GeV$^2$, 50\% of $Q^{2}F_{\pi}$ comes
from the regions with $b \leq 3.6$ GeV$^{-1}$
[$\alpha_s(1/b)/\pi \leq 0.32$] and with $b \leq 3.0$ GeV$^{-1}$
[$\alpha_s(1/b)/\pi \leq 0.21$], respectively. A liberal standard to judge
the self-consistency of the perturbative method would be that half of the
results arise from the region where $\alpha_s(1/b)/\pi$ is not larger than,
say, 0.5. From this point of view, PQCD becomes self-consistent at
about $Q^2\sim 4$ GeV$^2$. This conclusion is similar to that drawn in
\cite{LS}, where $\phi^{CZ}$ and $\phi^{AS}$ were employed. As
anticipated, the perturbative calculation improves with an increase of
the momentum transfer.

Results of $Q^2F_\pi$ from $\phi^{AS}$, $\phi^{CZ}$ and $\phi$ in
Eq.~(\ref{phie}), as well as the experimental data \cite{d1,d2,d3,CJB},
are shown in Fig.~5. The curve corresponding to Eq.~(\ref{phie}) is located
beyond the one corresponding to $\phi^{CZ}$ (and beyond the one
corresponding to $\phi^{AS}$, of course), indicating that the evolution of
the pion wave function, especially the portion from $1/b<\mu_0$, plays an
essential role. Our predictions for $Q^2 > 2$ GeV$^2$ are in a good
agreement with the data. For $Q^2 < 2$ GeV$^2$, the curve deviates from
the data, implying the dominance of the nonperturbative contrubution.
However, we emphasize that the wave function $\phi(x,1/b)$, regarded as an
expansion in $\alpha_s(1/b)/\alpha_s(\mu_0)$, becomes meaningless gradually
as $1/b\to \Lambda < \mu_0$. Hence, we shall not claim that the match of
our results with the data is conclusive. It is possible that
higher-twist contributions from other Fock states are important as argued
in \cite{CD}. Nonetheless, the resummed PQCD predictions are indeed
reliable at least near the high end of the accessible energy range, where
the curve from $\phi$ approaches that from $\phi^{CZ}$.

At last, we evaluate the pion form factor using the double-$b$
factorization formula in Eq.~(\ref{f5}), and the results are presented in
Fig.~6. Those derived from the single-$b$ formula in Eq.~(\ref{15}) and the
experimental data are also displayed for comparison. It is found that the
two curves are consistent with each other especially in the region with
$Q^2\sim 5$-10 GeV$^2$. The consistency hints that the approximation
of neglecting the transverse momenta carried by the virtual quarks makes
sense. Therefore, we shall employ the same approximation in order to
simplify the multidimensional integrals, when evaluating the hard
scattering amplitudes for the proton form factor.

We have stated that the function ${\bar{\cal P}}(x,b,1/b)$ in Eq.~(\ref{pb}) 
contains the intrinsic $b$ dependence denoted by its argument $b$, except
for the perturbative evolution with $b$ shown in Eq.~(\ref{phie}).
This intrinsic dependence, providing further suppression at large $b$, has
been discussed in detail in \cite{JK}. Since it depends on the 
parametrization, we shall not consider it here.

\section{The Perturbative Proton Form Factor}

With the success of the resummed PQCD formalism in the study of the pion
form factor, we are encouraged to apply it to the proton form factor. We
concentrate on the proton Dirac form factor $F_1^p$, which has been widely
calculated based on the standard PQCD theory
\cite{IL1,LB1,BR,JSL,CG,AET,BL,ACK,BI}. Another form factor
$F_2(Q^2)\sim O(1/Q^6)$ is much smaller than $F_1(Q^2)\sim O(1/Q^4)$
in the asymptotic region \cite{CZ1}. It has been shown that the asymptotic
proton wave function $\phi(x_i) \propto x_{1}x_{2}x_{3}$ fails to
give the correct sign for $F_{1}^{p}$, and generates much smaller values
than the experimental data \cite{AET,BI}. Highly asymmetric distribution
amplitudes, such as the Chernyak-Zhitnitsky (CZ) \cite{CZ1} and
King-Sachrajda (KS) \cite{KS} wave functions derived from QCD sum rules,
reverse the sign and enlarge the magnitude of the predictions. However, it
was also critically pointed out that the perturbative calculations based on the above
phenomenologically acceptable wave functions are typically dominated by
soft virtual gluon exchanges, and violate the assumption that the momentum
transfer proceeds perturbatively \cite{IL1}. We shall show that the
resummed PQCD formalism can give reliable predictions which are consistent
with the experimental data.

\subsection{Factorization Formula}

As stated in Sect.~3, the factorization formulas for exclusive QCD
processes require the investigation of those leading regions in the diagrams 
describing the radiative 
corrections where the dominant contributions arise from. 
Similarly, we work in the
axial gauge $n\cdot A=0$ as in the pion case. The two-particle reducible
diagrams, like Figs.~7(a) and 7(b), which are dominated by the collinear
divergences, are grouped into the distribution amplitudes, and generate
their Sudakov evolution. The bubbles may be regarded as the basic
photon-quark scattering diagrams. The two-particle irreducible diagrams
Figs.~7(c) and 7(d) contain only soft divergences, which cancel
asymptotically. These types of corrections are
then dominated by momenta of order $Q^2$, and absorbed into the hard
scattering amplitude $H$. Hence, the factorization formula for the proton
Dirac form factor includes three factors \cite{L}
\begin{eqnarray}
F_{1}^{p}(Q^{2})&=&\int_0^1 (d x)(d x')(d {\bf k}_T)(d {\bf k}'_T)
\bar{Y}_{\alpha'\beta'\gamma'}(k_{i}',P',\mu)
\nonumber \\
& &\times H_{\alpha'\beta'\gamma'\alpha\beta\gamma}(k_{i},k'_{i},Q,\mu)
Y_{\alpha\beta\gamma}(k_{i},P,\mu)\; ,
\label{3}
\end{eqnarray}
with $P=(P^+,0,{\bf 0})$ the initial-state proton momentum, and
\begin{eqnarray}
(d x)&=&d x_{1}d x_{2}d x_{3}\delta(\sum_{i=1}^{3}x_{i}-1)\;,
\nonumber\\
(d {\bf k}_T)&=&d^2{\bf k}_{1T}d^2{\bf k}_{2T}d^2{\bf k}_{3T}
\delta(\sum_{i=1}^{3}{\bf k}_{iT})\;.
\end{eqnarray}
$x_{i}=k_i^+/P^+$ and ${\bf k}_{iT}$ are the longitudinal momentum
fractions and transverse momenta of the parton $i$, respectively.
The primed variables $P'=(0,P^{'-},{\bf 0})$ and $x'_i=k_i^{'-}/P^{'-}$
are associated with the final-state proton. $Q^2=2P\cdot P'$
is the momumtum transfer, and the scale $\mu$ is the renormalization and
factorization scale. In the Breit frame we have $P^+=P^{'-}=Q/\sqrt{2}$.
Similarly, we have taken into account the transverse degrees of freedom of
a parton as indicated in Eq.~(\ref{3}). If the end-point regions are not
important, one may drop the $k_T$ dependence in the hard scattering $H$,
and integrate out $k_T$ in the proton distribution amplitude $Y$, arriving
at the standard factorization formula.

The initial distribution amplitude $Y_{\alpha\beta\gamma}$, defined by
the matrix element of three local operators in the axial gauge, is
written as \cite{BS,CZ1},
\begin{eqnarray}
Y_{\alpha\beta\gamma}&=&\frac{1}{2\sqrt{2}N_c}\int \prod_{l=1}^{2}
\frac{d y_{l}^{-}d{\bf y}_l}
{(2\pi)^{3}}e^{\textstyle ik_{l}\cdot y_{l}}\epsilon^{abc}
\langle 0|T[u_{\alpha}^{a}(y_{1})u_{\beta}^{b}
(y_{2})d_{\gamma}^{c}(0)]|P\rangle
\nonumber \\
&=&\frac{f_{N}(\mu)}{8\sqrt{2}N_{c}}
[(\,/\llap PC)_{\alpha\beta}(\gamma_{5}N)_{\gamma}V(k_{i},P,\mu)
+(\,/\llap P\gamma_{5}C)_{\alpha\beta}N_{\gamma}A(k_{i},P,\mu)
\nonumber \\
& &\mbox{ }-(\sigma_{\mu\nu}P^{\nu}C)_{\alpha\beta}(\gamma^{\mu}
\gamma_{5}N)_{\gamma}T(k_{i},P,\mu)]\; ,
\label{4}
\end{eqnarray}
where $|P\rangle$ is the initial proton state,
$u$ and $d$ the quark fields, $a$, $b$ and $c$ the color indices, and
$\alpha$, $\beta$ and $\gamma$ the spinor indices. In our notation, 1,2
label the two $u$ quarks and 3 labels the $d$ quark. The second form, 
showing the explicit Dirac matrix structure, is obtained from the spin
property of quark fields \cite{CZ1}, where $N$ is the proton spinor, $C$
the charge conjugation matrix and $\sigma_{\mu\nu}\equiv
[\gamma_{\mu},\gamma_{\nu}]/2$. The dimensional constant $f_{N}$ is
determined by the normalization of the distribution amplitude
\cite{I}. The amplitude $\bar{Y}_{\alpha'\beta'\gamma'}(k_{i}',P',\mu)$
for the final-state proton is defined similarly. The permutation symmetry
between the two $u$ quarks gives the relations among the wave functions
\cite{CZ1}
\begin{eqnarray}
& &V(k_1,k_2,k_3,P,\mu)=V(k_2,k_1,k_3,P,\mu)\;,
\nonumber \\
& &A(k_1,k_2,k_3,P,\mu)=-A(k_2,k_1,k_3,P,\mu)\;,
\nonumber \\
& &T(k_1,k_2,k_3,P,\mu)=T(k_2,k_1,k_3,P,\mu)\;.
\label{u1}
\end{eqnarray}
Since the total isospin of the three quarks is equal to $1/2$, we have
\cite{CZ1}
\begin{equation}
2T(k_{1},k_{2},k_{3},P,\mu)=\psi(k_{1},k_{3},k_{2},P,\mu)+
\psi(k_{2},k_{3},k_{1},P,\mu)\; , \label{5}
\label{u2}
\end{equation}
with
\begin{equation}
\psi(k_{1},k_{2},k_{3},P,\mu)=V(k_{1},k_{2},k_{3},P,\mu)-
A(k_{1},k_{2},k_{3},P,\mu)\; .
\label{u3}
\end{equation}
The combination of Eqs.~(\ref{u1}) and (\ref{u2}) leads to
\begin{eqnarray}
& &V(k_1,k_2,k_3,P,\mu)=\frac{1}{2}\;\left[\psi(k_2,k_1,k_3,P,\mu)
+\psi(k_1,k_2,k_3,P,\mu)\right]
\nonumber \\
& &A(k_1,k_2,k_3,P,\mu)=\frac{1}{2}\;\left[\psi(k_2,k_1,k_3,P,\mu)
-\psi(k_1,k_2,k_3,P,\mu)\right]\; .
\label{va}
\end{eqnarray}
Hence, the factorization formula in fact involves only one independent wave
function $\psi$.

The hard-scattering kernel $H_{\alpha'\beta'\gamma'\alpha\beta\gamma}$ can
be calculated perturbatively. To lowest order in $\alpha_{s}$ with two
hard exchanged gluons, 42 diagrams are drawn for the proton form factor.
This number is reduced to 21 when the permutation symmetry between the
incoming and outgoing protons is considered. These 21 diagrams are further
divided into two categories, which transform into one anoother by an 
interchange of the two $u$ quarks, so that it is enough to calculate only 11
of them, as shown in Fig.~8. The explicit expression of the integrand
$\bar{Y}_{\alpha'\beta'\gamma'}H_{\alpha'\beta'\gamma'\alpha\beta\gamma}
Y_{\alpha\beta\gamma}$, corresponding to each diagram in Fig.~8, is listed
in Table 1. For more details we refer to \cite{L}. The diagrams with
three-gluon vertices vanish at leading twist and are thus ignored.

Applying a series of variable changes \cite{L}, the summation of the
contributions over the 42 diagrams is reduced to only two terms. Equation
(\ref{3}) becomes
\begin{eqnarray}
F_{1}^{p}(Q^{2})&=&\frac{8\pi^{2}}{27}
\int_0^1 (d x)(d x')(d{\bf k}_{T})(d{\bf k}'_{T})[f_{N}(\mu)]^2
\nonumber \\
&\times&\sum_{j=1}^2\,H_j(k_i,k'_i,Q,\mu)\,\Psi_j(k_i,k'_i,P,P',\mu)\;.
\label{7}
\end{eqnarray}
with
\begin{eqnarray}
& &H_1=\frac{2\alpha_{s}^2(\mu)}
{3[(1-x_{1})(1-x_{1}')Q^{2}+({\bf k}_{1T}-{\bf k}'_{1T})^{2}]
[x_{2}x_{2}'Q^{2}+
({\bf k}_{2T}-{\bf k}'_{2T})^{2}]}\;,
\nonumber \\
& &H_2=\frac{2\alpha_{s}^2(\mu)}
{3[x_{1}x_{1}'Q^{2}+({\bf k}_{1T}-{\bf k}'_{1T})^{2}]
[x_{2}x_{2}'Q^{2}+({\bf k}_{2T}-{\bf k}'_{2T})^{2}]}\;,
\label{th}
\end{eqnarray}
and
\begin{eqnarray}
& &\Psi_{1}=\frac{2(\psi\psi')_{123}+8(TT')_{123}+2(\psi\psi')_{132}+
8(TT')_{132}-(\psi\psi')_{321}-(\psi\psi')_{231}}
{(1-x_{1})(1-x_{1}')}\;,
\nonumber  \\
& &\Psi_{2}=\frac{2(\psi\psi')_{132}-2(TT')_{123}}
{(1-x_{2})(1-x_{1}')}+
\frac{(\psi\psi')_{123}-8(TT')_{132}-2(\psi\psi')_{321}}
{(1-x_{3})(1-x_{1}')}\;.
\label{ps}
\end{eqnarray}
The notation $(\psi\psi')_{123}$ represents
\begin{equation}
(\psi\psi')_{123}=
\psi(k_1,k_2,k_3,P,\mu)\psi(k_1',k_2',k_3',P',\mu)\;,
\end{equation}
and $(TT')_{123}$ is defined in a similar way and we have inserted the 
electric charges of the quarks. 
The transverse momenta carried by the virtual quarks in
the hard scattering subdiagram have been neglected for simplicity as
indicated in Eq.~(\ref{th}). This approximation equates the initial and
final transverse distances among all pairs of valence quarks. At the same
time, fermion energies such as $x_iQ^2$ will not be among the
characteristic scales of the hard scattering, and the evolution of $H$ in
$Q$ will be simpler. The accuracy of this approximation has been roughly
discussed in Sect. 4.

\subsection{Sudakov Suppression}

Reexpressing Eq.~(\ref{7}) in the Fourier transform space, we obtain
\begin{eqnarray}
F_{1}^{p}(Q^{2})&=&\sum_{j=1}^2\frac{8\pi^{2}}{27}
\int_0^1 (d x)(d x')(d {\bf b})[f_N(\mu)]^2
\nonumber \\
& &\times {\tilde H}_j(x_i,x'_i,{\bf b}_{i},Q,\mu)
\Psi_{j}(x_{i},x'_i,{\bf b}_i,P,P',\mu)\; ,
\label{fp}
\end{eqnarray}
with ${\bf b}_{i}$ the conjugate variables to ${\bf k}_{iT}$, and
$(d{\bf b})=d^2 {\bf b}_1d^2 {\bf b}_2/(2\pi)^4$. The above factorization
formula involves only the integration over ${\bf b}$ (not ${\bf b}'$)
as stated above. The Sudakov resummation of the large logarithms in the
Fourier transformed wave function ${\cal P}$ leads to
\begin{eqnarray}
{\cal P}(x_i,{\bf b}_i,P,\mu)
&=&\exp\left[-\sum_{l=1}^3 s(x_lQ,w)-3\int_{w}^{\mu}
\frac{d\bar{\mu}}{\bar{\mu}}\gamma_{q}\left(\alpha_s(\bar{\mu})
\right)\right]
\nonumber \\
& &\times\phi(x_{i},w)\; ,
\label{8}
\end{eqnarray}
with the scale
\begin{equation}
w=\min(1/b_1,1/b_2,1/b_3)\;,
\label{ww}
\end{equation}
and $b_3=|{\bf b}_1-{\bf b}_2|$. The function $\phi$,
obtained by factoring the $Q$ dependence out of ${\cal P}$, corresponds to
the naive parton model. Again, we have dropped the intrinsic $b$
dependence of the proton wave function. The exponent $s(xQ,w)$ has been
given in Eq.~(\ref{fsl}).

Note the choice of the infrared cutoff in Eq.~(\ref{ww}), which differs from
that adopted in the previous analysis \cite{L}, but the same as 
$w=1/\max(b_1,b_2,b_3)$ in \cite{JKB}. In \cite{L} the different
transverse separations were assigned to each exponent $s$ and to each
integral involving $\gamma_q$:
\begin{eqnarray}
{\cal P}(x_i,{\bf b}_i,P,\mu)
&=&\exp\left[-\sum_{l=1}^3 \left(s(x_lQ,1/b_l)+
\int_{1/b_l}^{\mu}
\frac{d\bar{\mu}}{\bar{\mu}}\gamma_{q}\left(\alpha_s(\bar{\mu})
\right)\right)\right]
\nonumber \\
& &\times\phi(x_{i},w)\;.
\label{os}
\end{eqnarray}
The Sudakov factor in Eq.~(\ref{os}) does not suppress the soft
divergences coming from $b_l\to 1/\Lambda$ completely. For example, the
divergences from $b_1\to 1/\Lambda$, which appear in the integral involving
$\gamma_q$ and in $\phi(x_i,w)$ at $w\to \Lambda$, survive as $x_1\to 0$,
since $s(x_1Q,1/b_1)$ vanishes, while $s(x_2Q,1/b_2)$ and $s(x_3Q,1/b_3)$
may remain finite. On the other hand, $w$ should play the role of the
factorization scale, above which QCD corrections give the perturbative
evolution of ${\cal P}$ shown in Eq.~(\ref{8}), and below which they are
absorbed into the nonperturbative initial condition $\phi$. Hence, it is
not appropriate to choose the infrared cutoff of the Sudakov evolution
different from $w$. An alternative choice of the scale \cite{H},
\begin{equation}
w'=\max(1/b_1, 1/b_2, 1/b_3)\;,
\end{equation}
also removes the soft divergences. We argue that the factorization scale
should be chosen to diminish the largest logarithms contained in $\phi$,
which are proportional to $\ln(\mu/w)$. However, $w'$ does not serve this
purpose. In the numerical analysis below we shall truncate the integrations
over $b_1$ and $b_2$, in order to examine how much contributions
are accumulated in the short-distance (perturbative) region. Since small
values of $b_1$ and $b_2$ do not guarantee a small $b_3$ due to
the vector sum, we shall require $1/b_3$ to be greater than $\Lambda$, such
that the running coupling constant $\alpha_s(w)$ is well defined.

The evolution of the hard scattering amplitudes is written as
\begin{eqnarray}
{\tilde H}_j(x_{i},x_{i}',{\bf b}_i,Q,\mu)&=&
\exp\left[-3\sum_{l=1}^2\int^{t_{jl}}_{\mu}\frac{d\bar{\mu}}{\bar{\mu}}\,
\gamma_{q}\left(\alpha_s(\bar{\mu})\right)\right]
\nonumber \\
& &\times {\tilde H}_j(x_{i},x_{i}',{\bf b}_i,Q,t_{j1},t_{j2})\;,
\label{thj}
\end{eqnarray}
with
\begin{eqnarray}
& &t_{11}=\max\left[\sqrt{(1-x_{1})(1-x_{1}')}Q, 1/b_1\right]\;,
\nonumber \\
& &t_{21}=\max\left[\sqrt{x_{1}x_{1}'}Q, 1/b_1\right]\;,
\nonumber \\
& &t_{12}=t_{22}=\max\left[\sqrt{x_{2}x_{2}'}Q, 1/b_2\right]\; .
\label{tt}
\end{eqnarray}
The first scales in the brackets are associated with the longitudinal
momenta of the hard gluons and the second with the transverse
momenta. The arguments $t_{j1}$ and $t_{j2}$ of ${\tilde H}_j$ mean that
each $\alpha_s$ is evaluated at the largest mass scale of the corresponding
hard gluon.

Inserting Eqs.~(\ref{8}) and (\ref{thj}) into (\ref{fp}), we obtain
\begin{eqnarray}
F_{1}^{p}(Q^{2})&=&\sum_{j=1}^2\frac{4\pi}{27}
\int_0^1 (d x)(d x')\int_0^{\infty}
b_1 d b_1 b_2 d b_2 \int_0^{2\pi} d\theta [f_{N}(w)]^{2}
\nonumber \\
& &\times {\tilde H}_j(x_{i},x_{i}',b_i,Q,t_{j1},t_{j2})\,
\Psi_{j}(x_{i},x'_i,w)
\nonumber \\
& &\times \exp\left[-S(x_{i},x_{i}',w,Q,t_{j1},t_{j2})\right]\; ,
\label{10}
\end{eqnarray}
where
\begin{eqnarray*}
& &{\tilde H}_1=\frac{2}{3}\alpha_{s}(t_{11})\alpha_{s}(t_{12})
K_{0}\left(\sqrt{(1-x_{1})(1-x_{1}')}Q b_1\right)
K_{0}\left(\sqrt{x_{2}x_{2}'}Q b_2\right)
\nonumber \\
& &{\tilde H}_2=\frac{2}{3}\alpha_{s}(t_{21})\alpha_{s}(t_{22})
K_{0}\left(\sqrt{x_{1}x_{1}'}Q b_1\right)
K_{0}\left(\sqrt{x_{2}x_{2}'}Q b_2\right)
\label{k}
\end{eqnarray*}
are derived from the Fourier transform of Eq.~(\ref{th}), and the
expressions for $\Psi_j$ are the same as those in Eq.~(\ref{ps}) but with
$\psi(k_i,P,\mu)$ replaced by $\phi(x_i,w)$. The variable $\theta$ is
the angle between ${\bf b}_1$ and ${\bf b}_2$.
The Sudakov exponent $S$ is written as
\begin{eqnarray}
S(x_i,x'_i,w,Q,t_{j1},t_{j2})&=&\sum_{l=1}^3 s(x_lQ,w)+
3\int_w^{t_{j1}}\frac{d\bar{\mu}}{\bar{\mu}}
\gamma_{q}\left(\alpha_s(\bar{\mu})\right)
\nonumber \\
& &+\sum_{l=1}^3 s(x'_lQ,w)+
3\int_w^{t_{j2}}\frac{d\bar{\mu}}{\bar{\mu}}
\gamma_{q}\left(\alpha_s(\bar{\mu})\right)\;.
\label{ssp}
\end{eqnarray}
Similarly, the two-loop running coupling constant $\alpha_s$ in
Eq.~(\ref{2la}) is employed to derive the complete Sudakov factor.
This is an improvement of the analyses in \cite{L,JKB}.

The evolution of $f_{N}$ is given by
\begin{equation}
f_{N}(w)=f_{N}(\mu_{0})\left[\frac{\alpha_{s}(w)}{\alpha_{s}(\mu_{0})}
\right]^{1/(6\beta_0)}\; ,
\label{fn}
\end{equation}
with $\mu_{0}\approx 1$ GeV and the initial condition
$f_{N}(\mu_{0})=(5.2\pm 0.3)\times 10^{-3}$ GeV$^2$ \cite{CZ1}. For the
wave function $\phi$, we consider both the CZ model \cite{CZ1} and KS model
\cite{KS}. They are decomposed into
\begin{equation}
\phi(x_{i},w)=\phi^{AS}(x_{i})\sum_{j=0}^{5}N_{j}
\left[\frac{\alpha_{s}(w)}{\alpha_{s}(\mu_{0})}\right]^{b_{j}/(4\beta_0)}
a_{j}A_{j}(x_{i})\; ,
\label{wf}
\end{equation}
where the constants $N_{j}$, $a_{j}$ and $b_{j}$, and the first six Appel
polynomials $A_{j}(x_{i})$ \cite{LB1,BL} are listed in Table 2. The 
function $\phi^{AS}(x_{i})=120x_{1}x_{2}x_{3}$ is the asymptotic form of
$\phi$ mentioned before.

\subsection{Numerical Results}

If we employ the Sudakov exponent in Eq.~(\ref{ssp}) directly, we obtain
the results $Q^4F_1^p=0.55$ GeV$^4$ for $Q^2=10$ GeV$^2$ and
$Q^4F_1^p=0.48$ GeV$^4$ for $Q^2=40$ GeV$^2$ by using the KS wave function,
which are only about half of the experimental data. These values are close
to those derived in \cite{JKB}. However, we argue that there is a freedom to
choose the scales for the Sudakov evolution \cite{BS}. For example, the
ultraviolet cutoff $c\xi Q/\sqrt{2}$ for the integral in Eq.~(\ref{fsl}),
with $c$ a constant of order unity, is as appropriate as $\xi Q/\sqrt{2}$.
Hence, we may determine $c$ by fitting the predictions for the pion form
factor to the data $Q^2F_\pi\approx 0.35$ GeV$^2$ for $Q^2\sim 5$-10
GeV$^2$, and apply it to the evaluation of the proton form factor. This
procedure leads to $c\approx 0.6$. The change from $c=1.0$ to
$c=0.6$ enhances the results of the pion form factor by about 15\%. We
expect that this change in $c$ will give a larger enhancement in
the proton case, since the proton form factor is more sensitive to the
evolution scales. We emphasize that the parameter $c=0.6$ is introduced 
in order to explain the data. It is certainly legitimate to question this
smaller value of $c$, and argue that higher Fock state contributions may be
important.

In the numerical analysis below we adopt the Sudakov exponent
$S(x_i,x'_i,w,0.6Q,t_{j1},t_{j2})$.
We investigate the sensitivity of our predictions to the large
$b_1$-$b_2$ region, choosing $\Lambda=0.2$ GeV. If the perturbative
contributions dominate, an essential amount of $Q^4F_1^p$ will be quickly
accumulated below a common cutoff $b_{c}$ of $b_1$ and $b_2$. The numerical
outcomes are displayed in Fig.~9. All the curves, showing the dependence of
$Q^4F_1^p$ on $b_c$, increase from the origin and reach their full height
at $b_{c}=1/\Lambda$. A quantitative measure of self-consistency is that
half of the contribution to $Q^{4}F_{1}^{p}$ is accumulated in the region
characterized by $\alpha_s/\pi\le 0.5$. Based on this standard,
the results above $Q^2 \sim 20$ GeV$^2$ are reliable. At a higher energy
$Q^2=40$ GeV$^2$, the contribution from the perturbative region amounts
approximately to 65\% of the entire contribution. It implies that the
applicability of PQCD to the proton form factor at currently accessible
energy scale $Q^2\sim 30$ GeV$^2$ may be justified.

Note that the saturation of the proton form factor in the large $b$ region
is slower than that in the pion case, because the Sudakov suppression for
the former is more moderate compared to that for the latter. 
The proton form factor involves more $\gamma_q$, which enhance the
nonperturbative contributions, and the partons are softer, thereby
weakening the Sudakov suppression. The slower transition of the proton form
factor to PQCD as $Q^2$ increases can also be understood as follows. The
hard gluon exchange diagrams give leading contributions of order
$(\alpha_s/\pi)^2$, rather than $\alpha_s/\pi$ as in the pion case,
so that soft contributions, always suppressed by a power of $M^2/Q^2$ with
$M$ a typical hadronic mass, become relatively stronger \cite{R}.

In Fig.~10 we compare our predictions for $Q^{4}F_{1}^{p}(Q^{2})$ to
the experimental data, which are extracted from the data of elastic
electron-proton cross section \cite{JSL,A}. The match between the curve
corresponding to the KS wave function and the data is obvious. Below
$Q^2 \sim 20$ GeV$^2$, the deviation of the curve from the data becomes
larger. The CZ model generates results which are smaller than those from
the KS model by a factor 2/3. The above analysis implies that the KS wave
function is more phenomenologically appropriate. We shall adopt this model
for the proton wave function in the future studies. The sensitivity of our
predictions to the variation of the QCD scale parameter $\Lambda$ is also
examined. We find that the decrease of $\Lambda$ to 0.1 GeV enhances
$Q^4F_1^p$ by about 10\%.

\section{Sum Rules for Compton Scattering}

In this section we review the derivation
of the local duality sum rule for pion Compton scattering. 
The theoretical motivations and justifications of the validity 
of sum rules for the invariant amplitudes of RCS or VCS can be found in 
\cite{CRS}.
 
As emphasized by Nesterenko and Radyushkin in their work on the sum rule
analysis of the pion form factor \cite{NR1}, the dominant contribution
to the intermediate energy behaviour of the pion form factor comes from a 
``soft'' diagram, which is nonfactorizable and is interpreted as an overlap
of initial and final state wave functions. According to this viewpoint, the
inclusion of a hard scattering in the intermediate state, as suggested by
the PQCD factorization formalism, does not describe the correct dynamics.
There is at the moment no general consensus on this issue, and the
experimental input is going to be crucial in order to rule out some of the
alternatives.

We remind briefly that the sum rule analysis of VCS requires a detailed
study of the properties of analyticity of the diagrams appearing in the 
Operator Product Expansion of a 4-current correlator, and this
task has been performed in detail in \cite{CC2}. The existence of a finite
contour dispersion relation, which allows the application of the usual OPE
technique, has also been shown to hold with sufficient accuracy. The sum
rules obtained by this method, which are similar to the form factor sum
rules, are now characterized by both an ``$s$'' and a ``$t$'' dependence,
and are shown to hold at a fixed value of the ratio $-t/s$. The basic idea
of this approach is to connect the invariant amplitudes of a ``true''
({\it i.e.}, with on-shell pions or protons) scattering to an OPE series,
which is expressed in terms of a perturbative ``spectral function''
and of the corresponding condensates.

We show that the sum rule can be organized in a rather compact form in the
Breit frame of the pion, where the spectral density becomes
{\em polynomial} in a variable called $Q$. In the sum rule
description of both form factors and Compton scattering, $Q$ is a function
of various kinematic invariants, such as the pion virtualities (here
denoted as $s_1$, $s_2$), plus the usual $s$ and $t$ invariants. In fact,
we recall that the OPE side of the sum rule is calculated with off-shell
pions, since it has to correspond to a short-distance factorization.
This variable $Q$, which parametrizes the large light-cone components of
the incoming (or outgoing) hadron in the scattering process, is used
to analyze the power suppression of the various contributions to the
spectral function. $Q$ is then averaged in the dispersion integral over the 
virtualities of the hadron momenta. 

We also evaluate the power corrections proportional to the quark
and gluon condensates, and give the complete expressions of the spectral functions
for the helicity amplitudes involved in pion Compton scattering. A new
Borel sum rule is derived and compared to the corresponding one for the
pion form factor. The Borel transform is defined in its ``contour form'',
and is used to reduce the ``noise'' or ``background
contribution'' to the sum rule. Ultimately, the success of any sum rule
program relies on the choice of a ``good'' correlator with minimal noise
and maximum projection onto the physical states that we intend to study.
Assessing the error in this unobvious re-organization of the perturbative
expansion, via the OPE, is nontrivial and requires a case-by-case study. We
remind that sum rules are just a different model, closer to a PQCD
description, and like all models, should be handled with caution. 
In some situations they can help us describe the underlying dynamics
better than other models,  and, as such, can be a very useful tool of
analysis.

\subsection{Overview of the Method}

We start our discussion by quoting the main results of \cite{CRS}.
Consider the following 4-point amplitude as graphically described in
Fig.~11,
\begin{eqnarray}
\Gamma_{\nu\lambda\mu\sigma}(p_1^2,p_2^2,s,t)&=&i\int{\rm d}^4x\,
{\rm d}^4y\,{\rm d}^4z\exp(-ip_1\cdot x+ip_2\cdot y-iq_1\cdot z)
\nonumber \\
& &\times \langle 0|T\left(\eta_{\nu}(y)J_{\lambda}(z)J_{\mu}(0)
\eta_{\sigma}^{\dagger}(x)\right)|0\rangle \; ,
\label{tp}
\end{eqnarray}
where
\begin{eqnarray}
J_{\mu}=\frac{2}{3}\bar{u}\gamma_{\mu}u -\frac{1}{3}\bar{d}\gamma_{\mu}d\;,
\;\;\;\;\;\;
\eta_{\sigma}=\bar{u}\gamma_5\gamma_{\alpha}d\;,
\label{jd}
\end{eqnarray}
are the electromagnetic and axial currents, respectively, of the $u$ and
$d$ quarks. The choice of this specific correlator is motivated by the
particular process we are considering, namely, pion Compton scattering.
The pions of momenta $p_1$ and $p_2$ are assumed to be off-shell,
while the photons of momenta $q_1$ and $q_2$ are on-shell.

The invariant amplitudes $H_1$ and $H_2$ of pion Compton scattering are
extracted from the matrix element,
\begin{equation}
M_{\lambda\mu}(p_1,p_2,q_1)= i\int d^4y \, e^{-iq_1 \cdot y}
\langle p_2|T\left(J_{\nu}(y)J_{\lambda}(0)\right) |p_1 \rangle\;,
\label{mnula}
\end{equation}
by expressing it as \cite{CRS}
\begin{equation}
M^{\lambda\mu}(p_1,p_2,q_1)= H_1(s,t) e^{(1)\lambda}e^{(1)\mu} +
H_2(s,t) e^{(2)\lambda}e^{(2)\mu}\;,
\label{h1h2}
\end{equation}
where the polarization vectors $e^{(1)}$ and $e^{(2)}$ will be defined
below. $M_{\nu\lambda}$ appears as the residue of the double pion pole at
$p_1^2=p_2^2=m_\pi^2$ in the 4-point Green function
$\Gamma_{\nu\lambda\mu\sigma}$. It is convenient to define a double
discontinuity \cite{CRS}
$\Delta_{\nu\lambda\mu\sigma}$ of $\Gamma_{\nu\lambda\mu\sigma}$ across
the $p_1^2$ and $p_2^2$ cuts by
\begin{equation}
\Delta_{\nu\lambda\mu\sigma}
\equiv
\Gamma_{++} - \Gamma_{+-} - \Gamma_{-+} +\Gamma_{--}\ ,
\end{equation}
where we define $\Gamma_{++}=\Gamma_{\nu\lambda\mu\sigma}
(p_1^2+i\epsilon,p_2^2+
i\epsilon)$.  

The poles may be isolated from the $x_0 \to - \infty$, $y_0 \to \infty$
limit of $\Gamma_{\nu\lambda\mu\sigma}$ in Eq.~(\ref{tp}), by inserting
complete sets of states in the resulting time ordering:
\begin{eqnarray}
\Gamma_{\nu\lambda\mu\sigma} &=&
{i\over (2\pi)^6} \int d^4x d^4y d^4z\, d^4{p'}_1 d^4{p'}_2\,
\theta(y_0) \theta(-x_0)\nonumber \\
&\ & \times
e^{-ip_1\cdot x-iq_1\cdot z+ip_2\cdot y}
 \nonumber \\
&\ &\times\delta({p'}_1^2 -m_\pi^2)
\delta({{p'}_2}^2 - m_\pi^2) \langle0|\eta_{\sigma}(y)|p'_2\rangle
\nonumber \\
&\ &\times\langle p'_2|T(J_{\nu}(z)J_{\lambda}(0))|p'_1\rangle
\langle p'_1|\eta_{\mu}^\dagger(x)|0\rangle \ + \ \Gamma^{cont},
\end{eqnarray}
where $\Gamma^{cont}$ includes contributions from higher states and
other time orderings, that do not contribute to the double pion pole. 
We express the matrix elements of the axial currents as
\begin{equation}
\langle 0|\eta_{\sigma}(y)|p'_2\rangle=i f_\pi {p'_2}_{\sigma}
e^{-i{p'}_2\cdot y}\;,
\end{equation}
\begin{equation}
\langle p'_1|\eta_{\mu}^\dagger(x)|0\rangle =-i f_{\pi}
{{p'}_1}_{\mu}e^{i{p'}_1 \cdot x}\;,
\end{equation}
which relate the single pion states to the vacuum at times
$\pm \infty$. We then derive the relation
\begin{eqnarray}
\Delta_{\nu\lambda\mu\sigma}&=&
f_\pi^2\, p_{1\mu}p_{2\sigma}\, (2\pi)^2\delta(p_1^2-m_\pi^2)
\delta(p_2^2-m_\pi^2)
\nonumber \\
&\ &\hskip 0.4 true in \times\ M_{\nu\lambda}(p_1,p_2,q_1)+\Delta^{cont}\;,
\end{eqnarray}
with $\Delta^{cont}$ the double discontinuity of the continuum contribution
$\Gamma^{cont}$.

On one hand, the correlator in Eq.~(\ref{tp}) is expanded in terms of the
diagrams in Figs.~12. Specifically, Fig.~12 represents the perturbative
contribution to the expansion, and Fig.~13 describe the power
corrections, which are proportional to the condensates of quarks and
gluons, respectively. The power corrections parametrize the nonperturbative
structure of the QCD vacuum. When the two external photons are physical and
all the Mandelstam invariants are moderately large (see also the discussion
in \cite{CL}), the contribution to the leading scalar spectral function
comes from the diagrams in Fig.~12. These diagram, compared to the
corresponding one in the form factor case, contain an additional internal
line (the line that connects the two photons), which is far off-shell. We
can confirm that the process is, indeed, of hand-bag type when $s$ and $t$
are moderately large, with $s\gg t$. As discussed in \cite{CRS}, this
picture breaks down at small $t$, near the forward region. Note that
Fig.~12(a) does not contain any exchanged gluons, differently from
the expansion based on the PQCD factorization theorem \cite{CL}, where
the one-gluon-exchange diagrams are supposed to dominate. 

On the other hand, the resonance region is modeled, as usual, by a single
resonance (the double pole at the squared pion mass $m_\pi^2$) and a
continuum contribution starting at virtuality $s_0$. Information on the
invariant amplitudes of pion Compton scattering is obtained from the
residue at the double pion pole of the correlator in Eq.~(\ref{tp}).

We choose the pion Breit frame, in which the pion momenta are parametrized
as
\begin{equation}
p_1=Q \bar{v} +{s_1\over 2 Q} v\;, \,\,\,\,\,\,\,\,\,\,\, p_2={s_2\over 2
Q} \bar{v} + Q v\;.
\end{equation}
The dimensionless vectors $v$ and $\bar{v}$ are on the light cone, obeying
$v^2 = \bar{v}^2=0$ and $v\cdot\bar{v}=1$. The momentum transfer $t$ is
expressed in terms of $Q^2$, which acts as a large ``parameter" in the
scattering process, as
\begin{equation}
t=(p_2-p_1)^2=s_1+s_2 -2 Q^2- {s_1 s_2 \over 2 Q^2}.
\label{tq}
\end{equation}
Notice $t=-2 Q^2$ for $s_1=s_2=0$.
$Q^2$ is then given by
\begin{equation}
\label{qquadro}
Q^2={1\over 4} (s_1+s_2-t + \delta) = {1 \over 4} (s+u+\delta)\; ,
\end{equation}
with
\begin{equation}
\delta=\sqrt{(s_1+s_2-t)^2 -4 s_1 s_2} = {4 Q^4 - s_1 s_2\over 2 Q^2}\;.
\label{deltadef}
\end{equation}
In the frame specified above, all the transverse momenta are carried by the
two on-shell photons with physical polarizations. The advantage of using
such a frame is that the spectral densities are rational functions of
the variables $Q^2$, $s$, $s_1$ , and $s_2$. They are also 
symmetric in $s_1$ and $s_2$, as expected from the time reversal
invariance of the scattering process. This provides a drastic simplification
of the analysis. 

The matrix element in Eq.~(\ref{mnula}) admits various possible extensions
in terms of the polarization vectors, when the two pions go off shell. The
extension must be performed in such a way that the requirements of the
physical polarizations for the two photons,
\begin{eqnarray}
e^{(i)} \cdot q_1 &=& e^{(i)}\cdot q_2 = 0\;,
\nonumber \\
e^{(i)}\cdot e^{(j)} &=& -\delta_{ij}\;,
\label{eortho}
\end{eqnarray}
are still satisfied. A suitable choice of the polarization vectors will
simplify the calculations. The two vectors $e^{(1)}$ and
$e^{(2)}$ that we shall use differ slightly from those given in \cite{CRS}:
\begin{equation}
e^{(1)\lambda}={N^{\lambda}\over \sqrt{-N^2}}\;, \,\,\,\,\,\,\,\,
e^{(2)\lambda}={P^{\lambda}\over \sqrt{-P^2}}\;,
\label{edef}
\end{equation}
with
\begin{equation}
N^{\lambda}=\epsilon^{\lambda\mu\nu\rho}P_{\mu}r_{\nu}R_{\rho}\;,
\,\,\,\,\,\,\,\,
P^{\lambda}=\nu_1 p_1^\lambda+ \nu_2 p_2^\lambda +{\alpha\over2} R^\lambda\;,
\end{equation}
\begin{equation}
\nu_1 = {p_1\cdot p_2 - s_2}\;, \,\,\,\,\,\,\,\nu_2=p_1\cdot p_2-s_1\;,
\label{n12}
\end{equation}
\begin{equation}
\alpha=\frac{2}{t}(\nu_1 p_1\cdot R + \nu_2 p_2\cdot R)\;,
\end{equation}
\begin{equation}
R=q_1+q_2\;, \ \ \ \ \ \ \ \ r=q_2 -q_1\;.
\label{rq}
\end{equation}
In the Breit frame $e^{(1)}$ contains only transverse components
\begin{equation}
e^{(1)}=(0,0,e^{(1)}_{\perp})\;, \,\,\,\,\,\, \,\,\,\,
e^{(1)}_{\perp} \cdot q_\perp =0\;.
\end{equation}
At the pion pole, where $p_1^2=p_2^2=0$, one gets $\nu_1=\nu_2=1$.
With this extension, the polarization vectors $e^{(i)}$ still satisfty the
requirements in Eq.~(\ref{eortho}), which hold for all positive $s_1$ and
$s_2$, whether or not they are equal, and for all $\nu_1$ and $\nu_2$ as
specified in Eq.~(\ref{n12}).

The residue at the pion pole is isolated from the exact spectral density
of Eq.~(\ref{tp}) from a phenomenological ansatz of the form ``1 pole
plus continuum'', 
\begin{eqnarray}
\Delta_{\mu\nu\lambda\sigma}&=&
f_\pi^2\, p_{1\mu}p_{2\sigma}\, (2\pi)^2\delta(p_1^2-m_\pi^2)
\delta(p_2^2-m_\pi^2)\nonumber \\
&\ & \times M_{\nu\lambda}(p_1,p_2,q_1)+\Delta^{\rm pert} \,\,\,\,\,  +
{\rm h.\, r.}
\label{duv}
\end{eqnarray}
where $M_{\nu\lambda}$ has been defined in Eq.~(\ref{mnula}),
$\Delta^{\rm pert}$ is the double discontinuity of the multiparticle
states of the correlator in Eq.~(\ref{tp}), and ${\rm h.\,r.}$ denotes the
contributions from higher resonances. In the derivation of the sum rule it
is assumed that other single particle states, corresponding to the double
poles located at momenta $p_1^2$ and $p_2^2$ equal to the masses of higher
resonances, give negligible contributions. Therefore,
$\Delta_{\mu\nu\lambda\sigma}$ contains the most valuable nonperturbative
information on the invariant amplitudes $H_1$ and $H_2$ at the pion pole,
although mixed with the contributions of the higher states.

A suitable way to remove the contributions of single particle states from
Eq.~(\ref{duv}), but still keep the pionic contributions ($H_1$ and $H_2$),
is to use a projectior $n$, which satisfies the requirements \cite{CRS},
\begin{equation}
n\cdot q_1=n\cdot(p_1-p_2)=n^2=0\;.
\label{ndef}
\end{equation}
A possible choice is 
\begin{equation}
n^\mu = \biggl (e^{(2)} \pm ie^{(1)} \biggr )^\mu\; .
\label{nconstruct}
\end{equation}
The projector $n^\mu=(n^+,n^-,n_\perp)$ then has the components
\begin{eqnarray}
n^\pm &=& \left(\nu_1 {p_1}^\pm + {\nu_2}{p_2}^\pm
+{ \alpha\over 2}({q_1}^\pm+
{q_2}^\pm) \right)/\sqrt{-P^2}\;,
\nonumber \\
n_\perp &=&{\alpha q_\perp\over \sqrt{-P^2}} + i e^{(1)}.
\label{enne}
\end{eqnarray}
Note that only the transverse components of $n^{\mu}$ are imaginary.

For pion Compton scattering, the two projected spectral densities are
then given by the expressions,  
\begin{eqnarray}
\Delta^{(12)}(p_i^2,s,t)&=&
f_\pi^2\, n\cdot p_{1} n\cdot p_{2}\, (2\pi)^2\delta(p_1^2-m_\pi^2)
\delta(p_2^2-m_\pi^2)
\nonumber \\
& & \times \left [ H_1(s,t)+ H_2(s,t) \right ]\, +
\Delta^{(12){\rm pert}}\;,
\\
\Delta^{(1)}(p_i^2,s,t)&=&
f_\pi^2\, n\cdot p_{1} n\cdot p_{2}\, (2\pi)^2\delta(p_1^2-m_\pi^2)
\delta(p_2^2-m_\pi^2) \nonumber \\
& & \times  H_1(s,t)\, +  \Delta^{(1){\rm pert}}\;.
\end{eqnarray}

\subsection{Spectral Functions from Local Duality}

We present the complete evaluation of the lowest-order perturbative
spectral functions $\Delta^{(1){\rm pert}}$ and $\Delta^{(12){\rm pert}}$,
which appear in the sum rules for $H_1$ and $H_1+H_2$, respectively.
In a more complete analysis of pion Compton scattering an asymptotic
expansion of the lowest-order spectral functions is no longer sufficient.
While the dominant $O(1/Q^4)$ and the next-to-leading contributions
can be obtained by a simple power counting of the various integrals,
the $O(1/Q^8)$ contributions require a full-scale calculation of all the
terms from Fig.~12. This turns out to be a rather lengthy task. 

The double cut contribution  leading spectral functions $\Delta^{(1){\rm pert}}$ and
$\Delta^{(12){\rm pert}}$ are calculated in terms of the 3-cut integrals
\begin{equation}
\label{if}
I[f(k^2,k\cdot p_1,...)]=\int
d^4k \, f(k^2,k\cdot
p_1,...){\delta_+(k^2)\delta_+((p_1-k)^2)
\delta_+((p_2-k)^2)\over (p_1-k + q_1)^2}\;,
\end{equation}
\begin{equation}
\label{ipf}
I'[f(k^2,k\cdot
p_1,...)]=\int d^4k
f(k^2,k\cdot p_1,...)\delta_+(k^2)\delta_+((p_1-k)^2)\delta_+((p_2-k)^2)\;.
\end{equation}
The analogy to the form factor case is clearly due to the
fact that Eq.~(\ref{if}) approximates the total discontinuity along the
positive $s_1$ and $s_2$ axes by a "3-particle cut", with the top line of
Fig.~12 kept off-shell. The dominance of this discontinuity compared to
others can be easily proved. Other cuts are either subleading, relevant to
forward scattering -not considered in this approach-, or not in the
physical region. The evaluation of the integrals in Eqs.~(\ref{if}) and
(\ref{ipf}), though not trivial, are simplified in the Breit frame. 

The final expression for $\Delta^{(12){\rm pert}} $ is given by
\begin{eqnarray}
\Delta^{(12){\rm pert}} &=&{5\over 3 (2 \pi)^3}
 \left[ {8 Q^2 (s- 2 Q^2)R_{12}(Q^2,s,s_1,s_2)\over  (4
Q^4-s_1 s_2)^5  (2 Q^2 s - s_1 s_2)}\right.
 \nonumber \\
&\ & \quad \quad \quad \quad \left.+
{8 Q^2 (u- 2 Q^2)R_{12}(Q^2,u,s_1,s_2)\over (4 Q^4 -s_1 s_2)^5
(2 Q^2 u - s_1s_2)}\right] \;,
\label{depert12}
\end{eqnarray}
with
\begin{equation}
R_{12}(Q^2,s,s_1,s_2) = \sum_{n=0}^8 b_n(s,s_1,s_2) Q^{2n}\;,
\label{R12}
\end{equation}
a polynomial of $Q^2$. $R_{12}(Q^2,u,s_1,s_2)$ in the second term in the
brackets has a similar expression with the argument $s$ in Eq.~(\ref{R12})
replaced by $u$. This term corresponds to the crossed diagram
Fig.~12(b), where the two photon lines are interchanged.
The explicit expressions of the coefficients $b_n$ can be found in
\cite{CL}.

Using the relation
\begin{equation}
\label{pole}
(n\cdot p_1)( n\cdot p_2) = \frac{s (s+t)}{-t}\;,
\label{np}
\end{equation}
at the double pion pole, the expression of the sum rule becomes
\begin{equation}
f_\pi^2 \left [ H_1(s,t ) + H_2 (s,t)\right ] = \int_{0}^{s_0}ds_1
\int_{0}^{s_0}ds_2
\,\rho^{(12)pert}(s_1,s_2,s,t)\;,
\end{equation}
with the spectral density
\begin{equation}
\rho^{(12){\rm pert}}={-t\over s
(s+t)}\Delta^{(12)pert}(s_1,s_2,s,t)\;.
\end{equation}
The derivation of Eq.~(\ref{pole}) is referred to Appendix A.
By a simple power counting in $Q^2$, it is easy to observe that the
spectral function in Eq.~(\ref{depert12}) behaves as $1/t^2$ asymptotically,
suppressed by one more power of $t$, compared to the leading perturbative
predictions from the dimensional counting rules \cite{BL1}. This behaviour,
noticed in \cite{CRS}, is similar to that of the pion form factor derived from QCD sum rules \cite{NR1}.

The next point to address is the calculation of the power corrections 
\cite{CC1}. Since the top of the handbag diagram in Fig.~12 is far off-shell, the
contributions from the condensates are similar to those in the form factor
case. Therefore, our arguments about which soft contributions to be
inserted in the OPE are based on the similarity between the form factor and
Compton scattering. In the kinematic region with moderate $s$, $t$ and $u$,
and with fixed $-t/s$, all the operator insertions (condensates) which lower the
virtuality of the quark line connecting the two photons 
are not taken into account. In calculating
the power corrections, the general tensor operator
$G_{\mu\nu}G_{\alpha\beta}$ is projected onto its corresponding scalar
average $\langle G^2\rangle=\langle G^{a}_{\mu\nu}G^{a \mu\nu}\rangle$
by the relation
\begin{equation}
\langle 0|G_{\mu\nu}G_{\alpha\beta}|0\rangle ={1_c\over 72 }\langle 0|
G^{a}_{\rho\sigma}G^{a \rho\sigma}|0\rangle (g_{\mu\alpha}g_{\nu\beta}
-g_{\mu\beta}g_{\nu\alpha})\;.
\label{ave}
\end{equation}
with $1_c$ the $3\times 3$ unit color matrix and
$G_{\mu\nu}=G^{a}_{\mu\nu}\lambda^a/2$,
$\lambda^a$ being the Gell-Mann matrices. Equation (\ref{ave}) can also be
recast into the form
\begin{equation}
\langle 0|G^a_{\mu\nu}G^a_{\alpha\beta}|0\rangle ={1\over 96 }\langle 0|
G^{a}_{\rho\sigma}G^{a \rho\sigma}|0\rangle (g_{\mu\alpha}g_{\nu\beta}
-g_{\mu\beta}g_{\nu\alpha})\;,
\end{equation}
using the properties of the Gell-Mann matrices.

We work in the Fock-Schwinger gauge defined by 
\begin{equation}
z^\mu A_\mu(z)=0\;,
\end{equation}
and employ a simplified momentum-space method to evaluate the Wilson
coefficients \cite{mallik}. For massive quarks, the fermion propagator in
the external gluonic background is written as \cite{mallik}
\begin{eqnarray}
S(p) &=& {\not{p} +m\over p^2 - m^2}\,\, + \,\,{1\over 2} i
{(\gamma^{\alpha}\not{p}\gamma^{\beta} G_{\alpha \beta}
- m \gamma_{\alpha}G^{\alpha\beta}\gamma_{\beta})\over (p^2- m^2)^2}
\nonumber \\
&\ &  + {\pi^2 \langle G^2\rangle m\not{p}
(m +\not{p})\over (p^2 -m^2)^4}\;.
\label{propagator}
\end{eqnarray}
The above expression simplifies considerably in the massless limit,
where the third term corresponds to the insertion of two external gluon
lines vanishes, and the second term corresponds to a single gluonic
insertion on each fermion line survives. The gluonic power corrections
are then given by
\begin{equation}
\label{pcgluon}
\Delta^{(12){\rm gluon}}={20\alpha_s\over 27 \pi^2}{
Q^8\tau^{\rm gluon}(Q^2,s,s_1,s_2)\over
s (2 Q^2 -s_1)^4  (2 Q^2 -s_2)^4 }\;,
\end{equation}
with
\begin{eqnarray}
\tau^{\rm gluon}(Q^2,s,s_1,s_2) &=&
8 Q^6 -24 Q^4 s +14 Q^2 s^2 -2 s^3 +7  Q^4 s_1
\nonumber \\
& &  +2 Q^2 s s_1 -3 s^2 s_1 +17 Q^4 s_2 -6 Q^2 s s_2 -s^2 s_2\;.
\label{gluonic}
\end{eqnarray}
Equation (\ref{pcgluon}) shows that the power corrections for the new sum
rules are calculable systematically in full analogy to the form factor
case. The power-law fall off of such contributions are $O(1/Q^6)$,
suppressed by $1/Q^2$ compared to the perturbative spectral function.

The quark power corrections can also be calculated \cite{CC1}. The average
of the 4-quark operators is written as
\begin{equation}
\langle 0|\bar{\psi}^a_\alpha\bar{\psi}^b_{\beta}\psi^c_\gamma
\psi^d_\delta|0\rangle
={1\over 144}\langle (\bar{\psi}\psi)^2\rangle
(\delta^{ad}\delta^{bc}\delta_{\alpha\delta}\delta_{\beta\gamma}-
\delta^{ac}\delta^{bd}\delta_{\alpha\gamma}\delta_{\beta\delta})\;.
\end{equation}
Combining all possible insertions consistent with the handbag behaviour 
of Fig.~11 we derive
\begin{eqnarray}
\Delta^{(12){\rm quark}}(s_1,s_2,s,t)=
{10\pi \alpha_s\tau^{\rm quark}(Q^2,s_1,s_2)\over
27 Q^2 s s_1^3 (-2 Q^2 + s_1)^2 s_2^3 (-2 Q^2 + s_2)^2}\;.
\label{dequo}
\end{eqnarray}
Since the close form of the Borel transformed quark power corrections can
be obtained, as shown below, we do not present the
explicit expression of $\tau^{\rm quark}(Q^2,s_1,s_2)$ here.

Adding together Eqs.~(\ref{depert12}), (\ref{pcgluon}) and (\ref{dequo}),
the OPE of the spectral density is summarized as
\begin{eqnarray}
&&\rho^{(12)}(s_1,s_2,s,t)
\nonumber \\
&&= \left({-t\over s
(s+t)}\right)\Delta^{(12)}(s_1,s_2,s,t)\;,
\nonumber \\
&&= \left({-t\over s(s+t)}\right)
\bigg\{\Delta^{(12)pert}(s_1,s_2,s,t)\theta(s_0-s_1)\theta(s_0-s_2)
\nonumber \\
&& \quad +\alpha_s\langle G^2\rangle{20\over 27 \pi^2}
{Q^8\tau^{\rm {gluon}}(Q^2,s,s_1,s_2)\over s(2 Q^2 -s_1)^4 (2 Q^2 -s_2)^4 }
\nonumber \\
&& \quad +{2\over9}\langle (\bar{\psi}\psi)^2
\rangle  {10\pi\alpha_s\tau^{quark}(Q^2,s_1,s_2)\over
27 Q^2 s s_1^3 (-2 Q^2 + s_1)^2 s_2^3 (-2 Q^2 +s_2)^2}
\delta(s_1-p_1^2)\delta(s_2-p_2^2)\bigg\}\;.
\label{ro12}
\end{eqnarray}
Equation (\ref{ro12}), once integrated along the $(0,\lambda^2)$ cut in each
of the $s_1$ and $s_2$ planes, gives the sum rule
\begin{equation}
f_\pi^2\left[{ H_1(s,t) +H_2(s,t)\over p_1^2p_2^2}\right]
=\int_{0}^{\lambda^2}ds_1\int_{0}^{\lambda^2}ds_2{\rho^{(12)}(s_1,s_2,s,t)
\over(s_1-p_1^2)(s_2-p_2^2)},
\label{withdeno}
\end{equation}
where we have included all the contributions to the sum rule under a single
integral. Certainly, we still have to apply the Borel transform to
(\ref{withdeno}) in order to obtain the Borel sum rule \cite{coli}.
Here $\lambda$ defines the radius of the contour for the dispersion
relation as shown in Fig.~14, and also fixes the parameter of the Borel
transform.

After applying the Borel transform, Eq.~(\ref{withdeno}) becomes
\cite{CC1}
\begin{eqnarray}
&&{f_\pi}^2\left[H_1(s,t) +H_2(s,t)\right]=\left({-t\over s(s+t)}\right)
\nonumber \\
&&\times \bigg\{\int_{0}^{s_0}ds_1\int_{0}^{s_0}ds_2
{320 Q^{14}\tau^{\rm pert}(Q^2,s_1,s_2)\over
3 s (s-2 Q^2 )(4 Q^4 - s_1 s_2)^5 (-2Q^2 s + s_1 s_2)}
\nonumber \\
&&\quad + {20\over 27 \pi^2}\alpha_s\langle
G^2\rangle\int_{0}^{\lambda^2}ds_1\int_{0}^{\lambda^2}ds_2
{ Q^8\tau^{gluon}(Q^2,s,s_1,s_2)\over
s (2 Q^2 -s_1)^4  (2 Q^2 -s_2)^4 }e^{-(s_1+s_2)/M^2}\nonumber \\
&&\quad\quad\times \left(1- e^{-(\lambda^2-s_1)/M^2}\right)
\left(1- e^{-(\lambda^2-s_1)/M^2}\right)
\nonumber \\
&&\quad +\pi\alpha_s\langle (\bar{\psi}\psi)^2\rangle\left( F_0(M,s,t) +
F_1(M,s,t)e^{-\lambda^2/M^2}
+ F_2(M,s,t)e^{-2\lambda^2/M^2} \right) \bigg\}\;,
\label{suma}
\end{eqnarray}
with 
\begin{eqnarray}
&&F_0(M,s,t)\nonumber \\
&&={80\over 27}
{(-4 M^4 s - 8 M^2 s^2 -2 s^3 -2 M^2 s t - s^2 t + 2 M^2 t^2 +2 s
t^2 + t^3)\over M^4 t^2}\;, \nonumber \\
&&F_1(M,s,t)= {8\over 27} {(2 M^2 + s +t)(4 M^2 s + 2 s^2 - s t - t^2)
\over M^4 t^2}\;,\nonumber \\
&&F_2(M,s,t)= -{320\over 27}{s\over t^2}\;,
\end{eqnarray}
$M$ being the Borel mass.

We remind that Borel transforms, in a finite region of analiticity, 
should be replaced by their representation as contour integrals \cite{CRS}. 
A study of the stability analysis of this sum rule shows that $\lambda$
can be chosen to coincide with the duality interval $s_0$ \cite{coli}.
We compare the above formula with the sum rule for the pion form factor,
in which the integrals for the gluon and quark power corrections
can be explicitly performed \cite{NR1,JS}:
\begin{eqnarray}
&& f_\pi^2 F_\pi(Q^2)=\int_{0}^{s_0}ds_1\int_{0}^{s_0}ds_2\rho_{\pi 3}
(s_1,s_2,t)e^{-(s_1+s_2)/M^2}
\nonumber \\
&& +{\alpha_s \langle G^2\rangle\over 12\pi M^4} +
\langle\left(\bar{\psi}\psi\right)^2\rangle {208\over 81}\pi\alpha_s
\left(1 +{2\over 13}{Q^2\over M^2}\right)\;,
\label{piss}
\end{eqnarray}
with
\begin{eqnarray}
&&\rho_{\pi 3}(s_1,s_2,t)
\nonumber \\
&& \quad \quad ={3\over 2 \pi^4} t^2 \left[ \left ({d\over dt}\right )^2 +
{t\over 3} \left ({d\over dt}\right )^3 \right]
{1\over [(s_1 + s_2 - t)^2 - 4 s_1 s_2)]^{1/2}}\;.
\label{ropi}
\end{eqnarray}
Note that the power corrections in Eq.~(\ref{piss}) tend to grow rather
quickly with $Q^2$, and invalidate the sum rule at $Q^2$ beyond 4-5
GeV$^2$, while the perturbative contribution decreases fast as $1/Q^4$.
These behaviours set a tight constraint on the validity of the sum rule.
The asymptotic $Q^2$ dependence of the perturbative term in
Eq.~(\ref{suma}) is similar to that in Eq.~(\ref{piss}), but the power
corrections are more stable. This fact is due to the extra angular
dependence of Compton scattering. To have a more precise and definite
answer to the above issue, a stability analysis of the sum rule in
Eq.~(\ref{suma}) and a determination of the Borel mass $M$ are required. 
This study has been performed in \cite{coli}.

\section{Perturbative Pion CS: a Comparison with QCD Sum Rules}

In this section we apply the resummed PQCD formalism to
pion Compton scattering. We review the calculation of the invariant
amplitudes $H_1$ and $H_2$ \cite{CL}, and compare them to the sum
rule results obtained in the previous section. A comparison of 
perturbation theory and QCD sum rule prediscutions is presented. 

\subsection{Factorization Formulas}

The reasoning leading to the factorization of pion Compton scattering
is similar to that for the pion form factor. We consider first the
factorization formula for $H=H_1+H_2$, which includes the Sudakov
resummation of the higher-power effects at lower momentum transfers.
Basic diagrams for pion Compton scattering in the PQCD approach are shown
in Fig.~15, which differ from those in Fig.~12 by an extra exchanged gluon.
The contribution to the hard scattering amplitude from each diagram in
Fig.~15, obtained by contracting the two photon vertices with $-g^{\mu\nu}$,
is listed in Table 3. All the contributions can be grouped into two terms
using the permutation symmetry. We have
\begin{eqnarray}
H(s,t)&=& \sum_{l=1}^2 \int_{0}^{1}d x_{1}d x_{2}\,
\phi(x_{1})\phi(x_{2})\int_{0}^{\infty} bd b{\tilde H}_l(x_i,s,t,b)
\nonumber \\
& & \times \exp[-S(x_i,b,Q)]\;,
\label{fcs}
\end{eqnarray}
similar to Eq.~(\ref{15}). Here the pions are assumed to be on-shell,
$p_i^2=0$, and $t=-2Q^2$ is obtained from Eq.~(\ref{tq}) by setting $s_i$ to
zero. The Sudakov factor $\exp(-S)$ is the same as that associated with the
pion form factor in Eq.~(\ref{16}). The function $\phi$ is taken as the
CZ model \cite{CZ1} in Eq.~(\ref{cz}), since it gives the predictions
for the pion form factor which are close to those from the wave function
$\phi$ in Eq.~(\ref{phie}). In this way we simplify the analysis.

The Fourier transformed hard scattering amplitudes ${\tilde H}_l$
are given by
\begin{eqnarray}
{\tilde H}_1
&=&\frac{16\pi{\cal C}_F(e_u^2+e_d^2)\alpha_s(w_1)}{(1-x_1)(1-x_2)}
K_0\left(\sqrt{|r_1|}b\right)\left(\frac{[(1-x_1)t+u][(1-x_2)t+u]}{s^2}
\right.
\nonumber \\
& &\left.+\frac{[(1-x_1)t+s][(1-x_2)t+s]}{u^2}-4(1-x_2)\right)
\label{h1}
\end{eqnarray}
from the classes of Figs.~16(a)-16(c), and
\begin{eqnarray}
{\tilde H}_2
&=&32\pi{\cal C}_F e_u e_d\alpha_s(w_2)\left[\theta(-r_2)
K_0\left(\sqrt{|r_2|}b\right)
+\theta(r_2)\frac{i\pi}{2}H_0^{(1)}\left(\sqrt{r_2}b\right)\right]
\nonumber \\
&  & \times\left(\frac{1}{x_1(1-x_1)}-
\frac{(1+x_2-x_1 x_2)t^2+(1+x_2-x_1)ut}{x_2(1-x_{1})s^2}\right.
\nonumber \\
& &\;\;\;\;\left.+\frac{1}{x_2(1-x_2)}-
\frac{(1+x_1-x_1x_2)t^2+(1+x_1-x_2)st}{x_1(1-x_{2})u^2}\right)
\label{hh}
\end{eqnarray}
from the classes of Figs.~16(d) and 16(e), with
\begin{eqnarray}
r_1=x_1x_2t,\;\;\;\;\;\;r_2=x_{1}x_{2}t+x_1u+x_2s\; .
\end{eqnarray}
$K_{0}$ and $H_0^{(1)}$ in Eqs.~(\ref{h1}) and (\ref{hh}) are the
Bessel functions in the standard notation. The imaginary contribution
appears, because the exchanged gluons in Figs.~16(d) and 16(e) may
go to the mass shell ($r_2=0$). The argument $w_l$ of $\alpha_s$
is defined by the largest mass scale in the hard scattering,
\begin{equation}
w_1=\max\left(\sqrt{|r_1|},\frac{1}{b}\right),\;\;\;\;\;\;
w_2=\max\left(\sqrt{|r_2|},\frac{1}{b}\right)\; .
\label{w12}
\end{equation}
As long as $b$ is small, soft $r_l$  does not lead to large $\alpha_s$.
Therefore, the nonperturbative region in the modified factorization
is characterized by large $b$, where Sudakov suppression is strong.
Equation (\ref{fcs}), as a perturbative expression, is thus
relatively self-consistent compared to
the standard factorization. Since the singularity associated with
$r_2=0$ is not even suppressed by the pion wave function, Sudakov effects
are more crucial in Compton scattering \cite{MF} than in the case of
form factors.

Following the similar procedures, we derive the first invariant amplitude
$H_1$. The extraction of $H_1$ can be performed
by contracting the two photon vertices with $e^{(1)\mu}e^{(1)\nu}$.
The derivation of $H_1$ is much simpler than that of $H_2$, because
$e^{(1)}$ is orthogonal to all of the momenta $p_i$ and $q_i$. The
contribution to the hard scattering amplitude associated with $H_1$ from
each diagram in Fig.~16 is also listed in Table 3.
The modified perturbative expression for $H_1$ is given by
\begin{eqnarray}
H_1(s,t)&=& \sum_{l=1}^2 \int_{0}^{1}d x_{1}d x_{2}\,
\phi(x_{1})\phi(x_{2})
\int_{0}^{\infty} bd b
{\tilde H}^{(1)}_l(x_i,s,t,b)
\nonumber \\
& & \times \exp[-S(x_i,b,Q)]\;,
\label{ph1}
\end{eqnarray}
with
\begin{eqnarray}
{\tilde H}^{(1)}_1
=\frac{8\pi{\cal C}_F(e_u^2+e_d^2)\alpha_s(w_1)}{(1-x_1)(1-x_2)}
\left[\frac{u}{s}+\frac{s}{u}+4-2x-2y\right]
K_0\left(\sqrt{|r_1|}b\right)
\label{1h1}
\end{eqnarray}
from the classes of Figs.~16(a)-16(c), and
\begin{eqnarray}
{\tilde H}^{(1)}_2&=&16\pi{\cal C}_Fe_u e_d\alpha_s(w_2)
\left[\theta(-r_2)K_0\left(\sqrt{|r_2|}b\right)
+\theta(r_2)\frac{i\pi}{2}H_0^{(1)}\left(\sqrt{r_2}b\right)\right]
\nonumber\\
& &\times\left[\frac{1}{x_1(1-x_1)}+\frac{1}{x_2(1-x_2)}
-\frac{t}{x_2(1-x_1)s}-\frac{t}{x_1(1-x_2)u}\right]
\label{1hh}
\end{eqnarray}
from the classes of Figs.~16(d) and 16(e).

\subsection{Numerical Results}

Based on the sum rules and the resummed perturbative formulas for $H_1$
and $H_2$ of the previous sections, we compute the magnitudes of these two
helicities. Sum rule predictions are obtained from Eq.~(\ref{suma}) with
the substitution of $s_0=0.6$ and $M^2=4$ GeV$^2$ \cite{CL}. The resummed
PQCD formula for $H_1$ in Eq.~({\ref{ph1}) is evaluated numerically, and
$H_2$ is derived by $H_2=H-H_1$. Results of $H_1$ and $H_2$ at different
photon scattering angles $\theta^*$ are shown in Fig.~16 and 17,
respectively, in which $|H_i|$ denotes the magnitude of $H_i$. Note that
the sum rule predictions for $H_1$ are negative, and those for $H_2$ are
positive. It is observed that the sum rule results decrease more rapidly
with momentum transfer $|t|$ \cite{BF}, and have a weaker angular dependence
compared to the PQCD ones. The PQCD predictions are always larger than those
from sum rules at $\theta^*=50^o$ $(-t/s=0.6)$, and are always smaller at
$\theta^*=15^o$ $(-t/s=0.2)$ in the range $4 < |t| < 16$ GeV$^2$. It implies
that large-angle Compton scattering might be dominated by perturbative
dynamics. These two approaches overlap at $\theta^*=40^o$ $(-t/s=0.5)$ and
at $|t|=4$ GeV$^2$, showing the transition of pion Compton scattering to
PQCD. The transition scale is higher at smaller angles. 

With these results for $H_1$ and $H_2$, we compute the cross section of
pion Compton scattering. The expression for the differential cross
section of pion Compton scattering in the Breit frame is derived in the
Appendix B as 
\begin{equation}
\frac{d \sigma}{d\cos\theta^*}=\frac{|H_1|^2+|H_2|^2}{256\pi t}
\left(\frac{s-u}{s}\right)^3\;,
\label{csb}
\end{equation}
which yields the results exhibited in Fig.~18.
In the angular range we are investigating, the PQCD and sum rule methods
predict opposite dependence on $\theta^*$: the PQCD results increase, while
the sum rule results decrease, with the photon scattering angle. The reason
for this difference is that the increase of the amplitudes with $\theta^*$
is not sufficient to overcome the increase of the incident flux (see the
Appendix B). At a fixed angle, the differential cross section, similar to
$H_1$ and $H_2$, drops as $|t|$ grows. Again, the transition scale is
around 4 GeV$^2$ for $\theta^*=40^o$.

The differential cross section of pion Compton scattering from a polarized
photon has been analyzed based on the standard factorization formula in
\cite{MT}. The relevant scattering amplitudes and cross section can be
expressed in terms of $H_1$ and $H_2$ evaluated here \cite{CL}.
It was found that our predictions are consistent with those obtained in
\cite{MT}. However, in \cite{MT} the coupling constant $\alpha_s$ is
regarded as a phenomenological parameter and set to 0.3, while here we consider
the running of $\alpha_s$ due to the inclusion of radiative corrections,
and its cutoff is determined by the Sudakov suppression. Furthermore, our
perturbative calculation is self-consistent in the sense that short-distance
(small-$b$) contributions dominate.

In our analysis, which is restricted to lowest order in $\alpha_s$, the
sum rules for the helicities are real, and therefore, the issue related
to the perturbative and nonperturbative nature of the phases of
Compton scattering is not addressed. This issue needs a further study.

\section{Conclusion}

In this work we have recalculated the pion form factor using the PQCD
factorization theorem, which modifies the standard exclusive
theory \cite{LB1} by including the transverse degrees of freedom of a
parton and the Sudakov resummation. The large logarithmic corrections
at the end points of the parton momentum fractions are summed into
an exponential factor, which suppresses the nonperturbative contributions
from soft partons. With the help of the Sudakov suppression, PQCD enlarges its
range of applicability down to the energy scale of few GeV. Compared to the
previous analyses, we take into account the evolution of the wave function
and the more complete expression of the Sudakov factor derived from the
two-loop running coupling constant. The experimental data are then
explained.

By extending the resummed formalism to the proton Dirac form factor
and choosing an appropriate ultraviolet cutoff for the Sudakov evolution,
we have found that perturbation theory becomes self-consistent for $Q^2$
beyond 20 GeV$^2$, and our predictions from the KS wave function match the
experimental data. Compared to the standard leading-order
and leading-power descriptions \cite{LB1,ER,FJ}, where the transition
to the perturbative behaviour is relatively slow, and the predictions are not
reliable even at the highest available energies \cite{IL1}, our approach
provides an improvement of the theory for QCD exclusive processes. In the
present calculation we have modified the infrared cutoff for the Sudakov
evolution, and render it the same as the factorization scale for the proton
wave function. By this modification, the soft divergences from the
factorization scale close to $\Lambda_{\rm QCD}$ are completely removed
by the Sudakov suppression. Our analysis shows that the KS wave function is
a better choice compared to the CZ wave function, since the latter gives
results which amount only to 2/3 of the data.

We have reviewed the previous works on the local duality sum rule for pion
Compton scattering and explained the derivation of the leading spectral
function for the two invariant amplitudes. 
The leading nonperturbative corrections to
the sum rules can also be calculated in a systematic way. 
It has been shown that the sum rules for pion Compton
scattering, though more complicated, has the same power-law fall-off
as that of the pion form factor in the related sum rule analysis \cite{NR1}.  

The application of the resummed PQCD formalism to pion Compton scattering
is also reviewed. We have given the modified perturbative expressions for
the invariant helicity amplitudes. Comparing the predictions for pion
Compton scattering from the resummed PQCD formalism with those from the sum
rule analysis, we observe clear overlaps between the two approaches.
The numerical results show that the transition to PQCD appears at
$|t|=4$ GeV$^2$ and at $\theta^*=40^o$ in the Breit frame. This suggests
that sum rule and PQCD methods are complementary tools in the
description of exclusive reactions, and the power-law fall-off of the
relevant amplitudes helps us locate the transition region.

A more convincing justification of our conclusion will come from the study
of other more complicated processes, such as proton Compton scattering
\cite{FZK}. We have started a complete analysis of VCS using the resummed
perturbation theory and QCD sum rules, and the results will be published
in the near future. Finally, we expect that our methods will find
applications in the entire class of elastic hadron-photon interactions at
intermediate energies.

\centerline{\bf Acknowledgements}
We thank A. Radyushkin and G. Sterman for discussions. 
C.C. thanks the Theory Group at the Institute for Fundamental Theory 
at the University of Florida at Gainesville 
and in particular Prof. P. Ramond for hospitality.  
The work of C. Savkli is supported under the DOE grant No. DE-FG02-97ER41032.

\renewcommand{\theequation}{A.\arabic{equation}}
\setcounter{equation}{0}
\vskip 1cm \noindent
\section{Appendix A. Kinematics}
\vskip 3mm \noindent

Let $q_1$ and $q_2$ be the momenta of the incoming and outgoing photons,
respectively, which are assumed to be on-shell ($q_1^2=q_2^2=0$).
Let $p_1$ and $p_2$ be the momenta of the incoming and outgoing pions.
The external pion states are off-shell and are characterized by the
invariants $s_1=p_1^2$ and $s_2=p_2^2$.
We define the Mandelstam invariants
\begin{equation}
s=(p_1+q_1)^2\;, \,\,\,\,\,\,\,\,\,\, t=(q_2-q_1)^2\;,
\,\,\,\,\,\,\,\,\,,\,\,
u=(p_2-q_1)^2\;,
\end{equation}
which satisfy the relation
\begin{equation}
s+t+u=s_1+s_2.
\end{equation}
It is also convenient to introduce the light-cone variables for the photon
momenta:
\begin{eqnarray}
& &q_1 = q_1^+ \bar{v} +q_1^- v +q_{1\perp}\;,
\nonumber\\
& &\bar{v}={1\over\sqrt{2}} (1,1,{\bf 0}_\perp)\;, \quad \quad \quad
v={1\over\sqrt{2}} (1,-1,{\bf 0}_\perp)\;,
\nonumber \\
& & q_{1\perp}\cdot n^{\pm} = 0\;,
\end{eqnarray}
with the convention
\begin{equation}
v^+= {1\over \sqrt{2}} (v^0 + v^3)\;, \,\,\,\,\,\,
\bar{v}={1\over \sqrt{2} }(v^0 -v^3)\;.
\end{equation}
Covariant expressions for $q_1^\pm$ are found to be
\begin{equation}
q_1^+={(s-2 Q^2)(2 Q^2-s_2)\over 2Q\delta},
\end{equation}
\begin{equation}
q_1^-={(2 Q^2-s_1)(2 Q^2 s-s_1 s_2)\over 4Q^3\delta}.
\end{equation}
In the Breit frame of the incoming pion we easily get
\begin{equation}
u=(p_2-q_1)^2=2 Q^2-s +{{s_1 s_2}\over 2 Q^2}.
\end{equation}

As discussed in Sect. 6, a crucial step in the derivation of the sum rules
is the appropriate choice of the projector $n^{\lambda}$.
Its real part is parametrized as
\begin{equation}
n_R= (n^+,n^-, {\alpha q_\perp\over \sqrt{-P^2}})\;.
\end{equation}
We easily obtain the expressions
\begin{equation}
\alpha={(4 Q^4 - 4 Q^2 s + s_1 s_2)\over 2 Q^2}\;,
\end{equation}
\begin{equation}
P^2={(2 Q^2 -s)(2 Q^2 - s_1)(2 Q^2 -s_2)(2 Q^2 s - s_1 s_2)\over 4 Q^4}\;.
\end{equation}
Note that both $P^2$ and $\alpha$ are negative. The relation
\begin{eqnarray}
n\cdot p_1 &=&\left(1/\sqrt{-P^2}\right) (n^+ p_1^- + n_- p_1^+)\;,
\\ \nonumber
&=& {(s- 2 Q^2)(-2 Q^2 s + s_1 s_2)\over 2 Q^2 }\;,
\end{eqnarray}
together with the condition $n\cdot p_1=n\cdot p_2$, lead to
\begin{equation}
{1\over n\cdot p_1 n\cdot p_2}= {(2 Q^2-s_1)(2 Q^2-s_2)\over
(2 Q^2-s)(2 Q^2 s- s_1s_2)}\;.
\end{equation}
At the pion pole ($s_1=s_2=0, Q^2=-t/2$), we have
\begin{equation}
{1\over n\cdot p_1 n\cdot p_2}= {-t\over s (s+t)},
\end{equation}
which is the result quoted in Eq.~(\ref{pole}).

\renewcommand{\theequation}{B.\arabic{equation}}
\setcounter{equation}{0}
\vskip 1cm \noindent
\section{Appendix B. The Scattering Amplitudes and Cross Section}
\vskip 3mm \noindent

In this appendix we derive the expressions for the amplitudes
of pion Compton scattering
and the cross section appearing in Sect 7. The differential cross section
of Compton scattering is defined by
\begin{equation}
d\sigma=\frac{|{\cal M}|^2}{F}d Q\;,
\end{equation}
where ${\cal M}$ is the scattering amplitude, $F$ is the incident flux,
and $\d Q$ is the phase space of the final states,
\begin{equation}
d Q=(2\pi)^4\delta^{(4)}(p_1+q_1-p_2-q_2)
\frac{d {\rm p}_2}{(2\pi)^3 p_2^0}
\frac{d {\rm q}_2}{(2\pi)^3 q_2^0}\;.
\label{phs}
\end{equation}
$F$ is defined by
\begin{equation}
F=|{\bf v}_{p_1}-{\bf v}_{q_1}|2p_1^0 2q_1^0=4p\cdot q=2s\;,
\label{inf}
\end{equation}
with ${\bf v}_{p}={\bf p}/p^0$ the velocity of the incoming particle.
It is easy to observe that $F$ increases with the photon scattering angle
in the Breit frame. Combining Eqs.~(\ref{phs}) and (\ref{inf}), we obtain
the general expression
\begin{equation}
d\sigma=\frac{|{\cal M}|^2}{32\pi s}d \cos\theta
\label{csc1}
\end{equation}
with $\theta$ the center-of-mass scattering angle.

Substituting $\cos\theta=(t-u)/s$ into eq.~(\ref{csc1}), we have the
expression which is invariant in both of the center-of-mass and Breit
frames. Then Eq.~(\ref{csc1}) can be easily converted into the one
in the Breit frame using the relation $\sin\theta^*/2=-t/(s-u)$ with
$\theta^*$ as defined before. We have
\begin{equation}
\frac{d \sigma}{d\cos\theta^*}=\frac{|{\cal M}|^2}{128\pi t}
\left(\frac{s-u}{s}\right)^3\;,
\end{equation}
where the scattering amplitude ${\cal M}$ is given by
\begin{equation}
{\cal M}=M_{\mu\nu}\epsilon_{1T}^{\mu}\epsilon_{2T}^{*\nu}\;,
\end{equation}
with $M_{\mu\nu}$ defined by Eq.~(\ref{h1h2}), and $\epsilon_T$
the polarization vector of the photon in the state $T$.
Inserting $|{\cal M}|^2=(|H_1|^2+|H_2|^2)/2$ for
pion Compton scattering from an unpolarized photon into the above
formula, we obtain Eq.~(\ref{csb}).

\newpage
Table 1. The expressions of the integrand $\bar{Y}_{\alpha'\beta'\gamma'}
H_{\alpha'\beta'\gamma'\alpha\beta\gamma}Y_{\alpha\beta\gamma}$
for the diagrams in Fig.~8. $\psi_{123}$
is the brief symbol for $\psi(k_{1},k_{2},k_{3},P,\mu)$
and $\psi'_{123}$ for
$\psi(k_{1}',k_{2}',k_{3}',P',\mu)$.

\[ \begin{array}{rc}   \hline\hline\\
{\rm Diagram} & \bar{Y}HY/(4\pi^{2}\alpha_{s}^{2}f_{N}^2/27) \\
\hline   \\
({\rm a})     &{\displaystyle
\frac{e_{u}(\psi_{123}\psi'_{123}+4T_{123}T'_{123})}
{(1-x_{1})(1-x_{1}')[(1-x_{1})(1-x_{1}')Q^{2}+({\bf k}_{T_1}
-{\bf k}'_{T_1})^{2}]
[x_{3}x_{3}'Q^{2}+({\bf k}_{T_3}-{\bf k}'_{T_3})^{2}]} } \\
        &    \\
({\rm b})   &  {\displaystyle
\frac{e_{u}(\psi_{123}\psi'_{123}+4T_{123}T'_{123})}
{(1-x_{1})(1-x_{1}')[(1-x_{1})(1-x_{1}')Q^{2}+({\bf k}_{T_1}
-{\bf k}'_{T_1})^{2}]
[x_{2}x_{2}'Q^{2}+({\bf k}_{T_2}-{\bf k}'_{T_2})^{2}]}} \\
        &   \\
({\rm c})   &  {\displaystyle
-\frac{e_{u}4T_{123}T'_{123}}
{(1-x_{1})(1-x_{2}')[x_{2}x_{2}'Q^{2}+({\bf k}_{T_2}-{\bf k}'_{T_2})^{2}]
[x_{3}x_{3}'Q^{2}+({\bf k}_{T_3}-{\bf k}'_{T_3})^{2}]} } \\
        &    \\
({\rm d})   & {\displaystyle
-\frac{e_{u}\psi_{123}\psi'_{123}}
{(1-x_{1})(1-x_{3}')[x_{2}x_{2}'Q^{2}+({\bf k}_{T_2}-{\bf k}'_{T_2})^{2}]
[x_{3}x_{3}'Q^{2}+({\bf k}_{T_3}-{\bf k}'_{T_3})^{2}]} }\\
        &   \\
({\rm e})   &   0\\
        &   \\
({\rm f})   &   0 \\
        &    \\
({\rm g})   &  {\displaystyle
\frac{e_{u}\psi_{213}\psi'_{213}}
{(1-x_{3})(1-x_{2}')[x_{2}x_{2}'Q^{2}+({\bf k}_{T_2}-{\bf k}'_{T_2})^{2}]
[x_{3}x_{3}'Q^{2}+({\bf k}_{T_3}-{\bf k}'_{T_3})^{2}]} }\\
        &     \\
({\rm h})   &  {\displaystyle
\frac{e_{d}(\psi_{123}\psi'_{123}+\psi_{213}\psi'_{213})}
{(1-x_{3})(1-x_{3}')[x_{2}x_{2}'Q^{2}+({\bf k}_{T_2}-{\bf k}'_{T_2})^{2}]
[(1-x_{3})(1-x_{3}')Q^{2}+({\bf k}_{T_3}-{\bf k}'_{T_3})^{2}]} }\\
        &   \\
({\rm i})   &  {\displaystyle
-\frac{e_{d}\psi_{123}\psi'_{123}}
{(1-x_{3})(1-x_{1}')[x_{1}x_{1}'Q^{2}+({\bf k}_{T_1}-{\bf k}'_{T_1})^{2}]
[x_{2}x_{2}'Q^{2}+({\bf k}_{T_2}-{\bf k}'_{T_2})^{2}]} }\\
        &   \\
({\rm j})  &   0 \\
        &    \\
({\rm k})  &  {\displaystyle
\frac{e_{d}4T_{123}T'_{123}}
{(1-x_{2})(1-x_{1}')[x_{1}x_{1}'Q^{2}+({\bf k}_{T_1}-{\bf k}'_{T_1})^{2}]
[x_{2}x_{2}'Q^{2}+({\bf k}_{T_2}-{\bf k}'_{T_2})^{2}]} }\\
\vspace{0.2cm}\\
\hline\hline
\end{array}   \]

\newpage

Table 2. The constants and the Appel polynomials for the proton
wave function $\phi(x_{i},w)$ of the CZ and KS
models \cite{CZ1,KS} in Eq.~(\ref{wf}).

\[ \begin{array}{cllccl} \hline\hline
j & a_{j}({\rm CZ}) &a_j({\rm KS})  & N_{j} & b_{j} & A_{j}(x_{i}) \\ \hline
0 & 1.00      &1.00     & 1     & 0     & 1  \\
1 & 0.410     &0.310    & 21/2  & 20/9  & x_{1}-x_{3} \\
2 & -0.550    &-0.370   & 7/2   & 24/9  & 2-3(x_{1}+x_{3}) \\
3 & 0.357     &0.630    & 63/10 & 32/9  & 2-7(x_{1}+x_{3})+8(x_{1}^{2}
                                          +x_{3}^{2})+4x_{1}x_{3} \\
4 & -0.0122   &0.00333  & 567/2 & 40/9  & x_{1}-x_{3}-(4/3)(x_{1}^{2}
                                          -x_{3}^{2}) \\
5 & 0.00106   &0.0632   & 81/5  & 42/9  & 2-7(x_{1}+x_{3})
                                          +14x_{1}x_{3}  \\
  &           &         &       & & +(14/3)(x_{1}^{2}+ x_{3}^{2})\\
\hline\hline
\end{array}  \]

\newpage
Table 3. The expressions of the hard scattering amplitudes $H$ and
$H^{(1)}$ corresponding to the diagrams in Fig.~16. Here we define
\begin{eqnarray}
& &D_1=x_1x_2t-({\bf k}_{T_1}-{\bf k}_{T_2})^{2}\;,\nonumber \\
& &D_2=x_{1}x_{2}t+x_1u+x_2s-({\bf k}_{T_1}-{\bf k}_{T_2})^{2}\;.
\nonumber
\end{eqnarray}

\[ \begin{array}{lcc}   \hline\hline\\
{\rm Diagram}& H/(16\pi\alpha_{s}{\cal C}_F) & H^{(1)}/(8\pi\alpha_{s}
{\cal C}_F) \\
\hline   \\
({\rm a})     &{\displaystyle
\frac{-e_{u}^2[(1-x_1)t+u][(1-x_2)t+u]}
{(1-x_1)(1-x_2)s^2D_1}}  &{\displaystyle
\frac{-e_{u}^2u}
{(1-x_1)(1-x_2)sD_1} }\\
        &  &   \\
({\rm b})   &  {\displaystyle
\frac{e_{u}^2}
{(1-x_{1})D_1}} & {\displaystyle
\frac{-e_{u}^2}
{(1-x_{1})D_1}}  \\
        &  &  \\
({\rm c})   &  {\displaystyle
\frac{e_{u}^2}
{(1-x_{1})D_1}} & {\displaystyle
\frac{-e_{u}^2}
{(1-x_{1})D_1}}  \\
        &   & \\
({\rm d})   &  {\displaystyle
\frac{-e_{u}e_d}
{x_1(1-x_{1})D_2}}
&   {\displaystyle
\frac{-e_{u}e_d}
{x_1(1-x_{1})D_2}} \\
        &  &  \\
({\rm e})   & {\displaystyle
\frac{e_{u}e_d[(1+x_2-x_1x_2)t^2+(1+x_2-x_1)ut]}
{x_2(1-x_{1})s^2D_2}}
 & {\displaystyle
\frac{e_{u}e_dt}
{x_2(1-x_{1})sD_2} }
\vspace{0.2cm}\\
\hline\hline
\end{array}   \]


\begin{figure}[ht]

\includegraphics[width=5in]{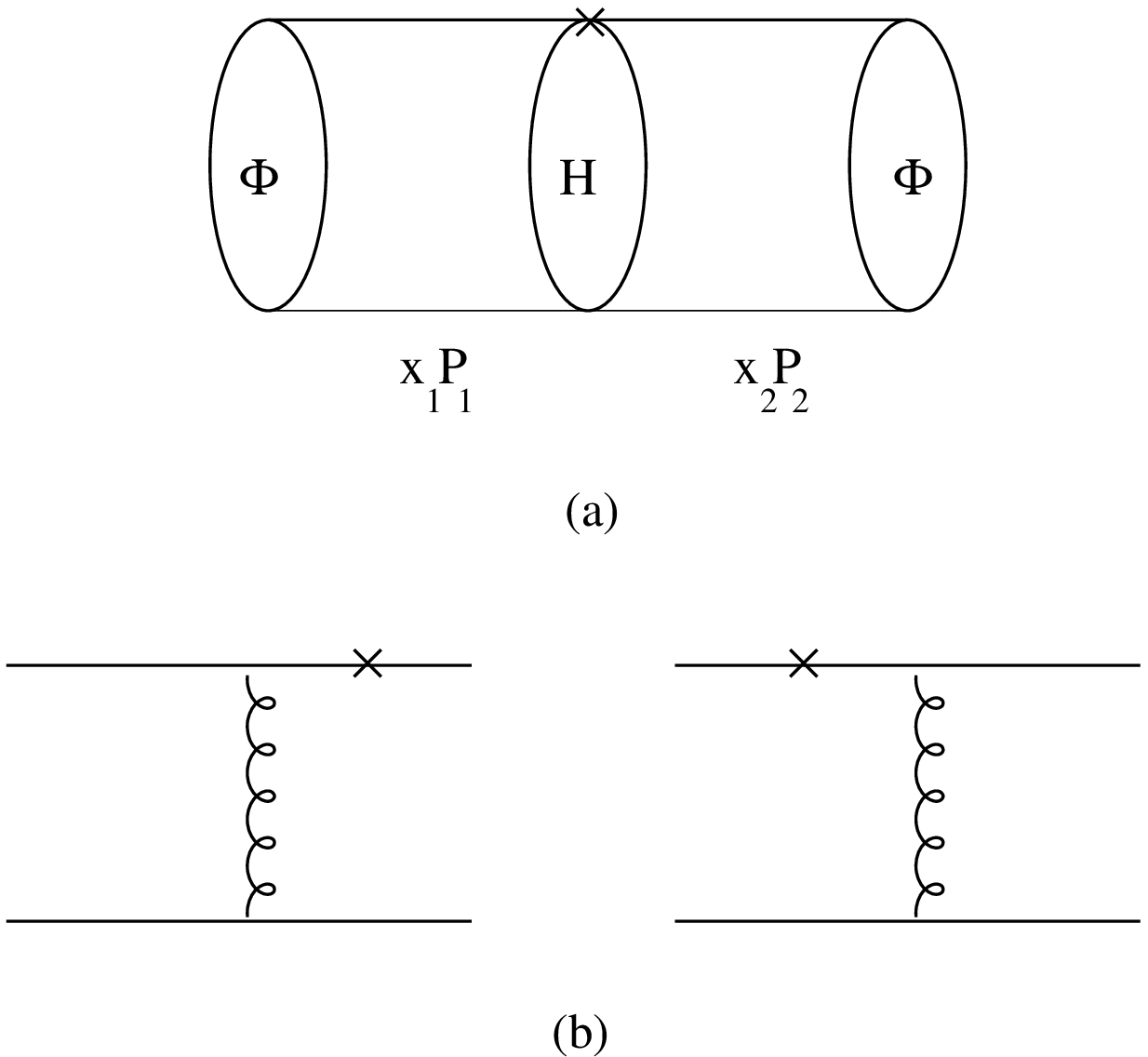}
\vspace{.2in}
\caption{(a) Factorization of the pion form factor. The symbol
$\times$ represents the photon vertex. (b) Basic scattering diagrams.}
\vspace{.2in}
\includegraphics[width=5in]{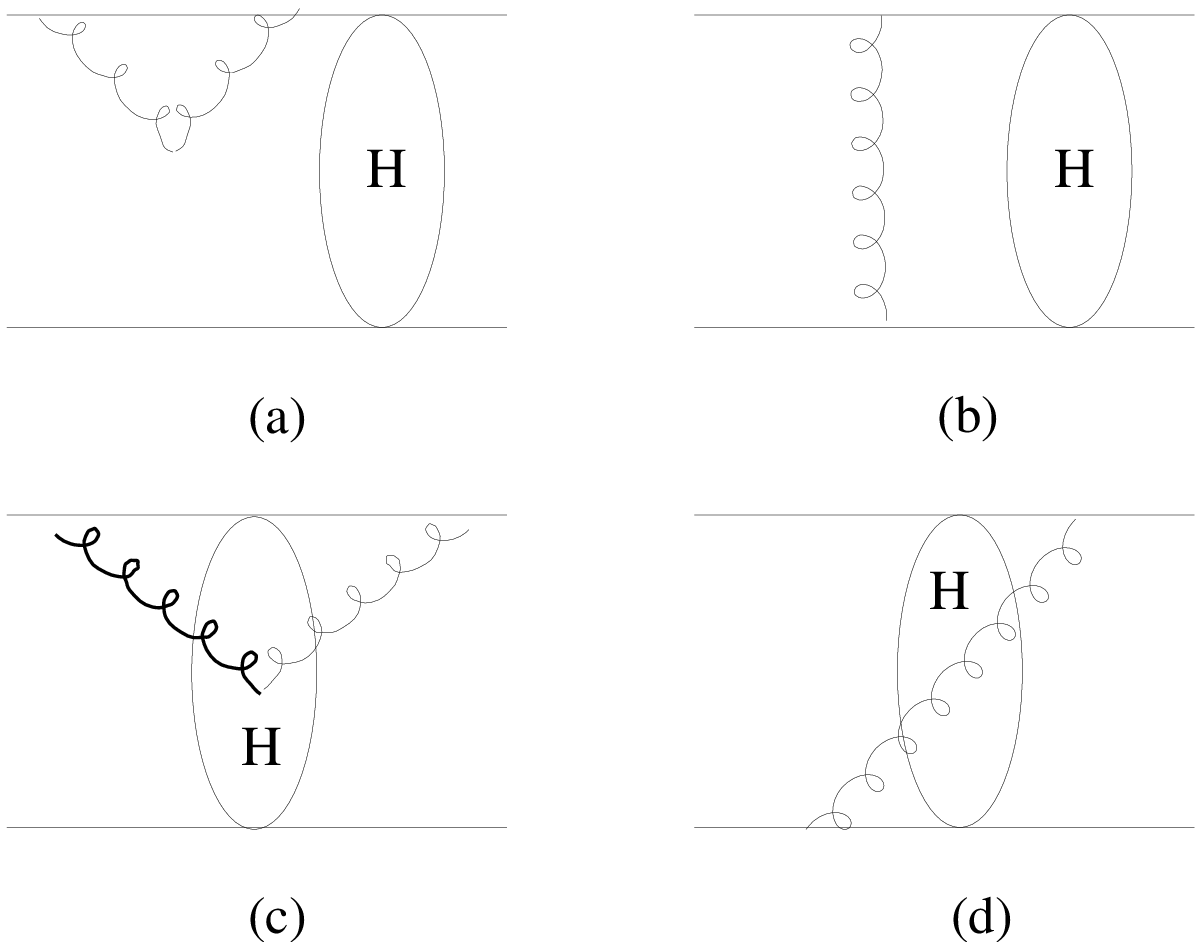}
\caption{Radiative corrections to the basic scattering diagrams in
Fig.~1(b).}
\end{figure}

\newpage
\begin{figure}[ht]
\includegraphics[width=5in]{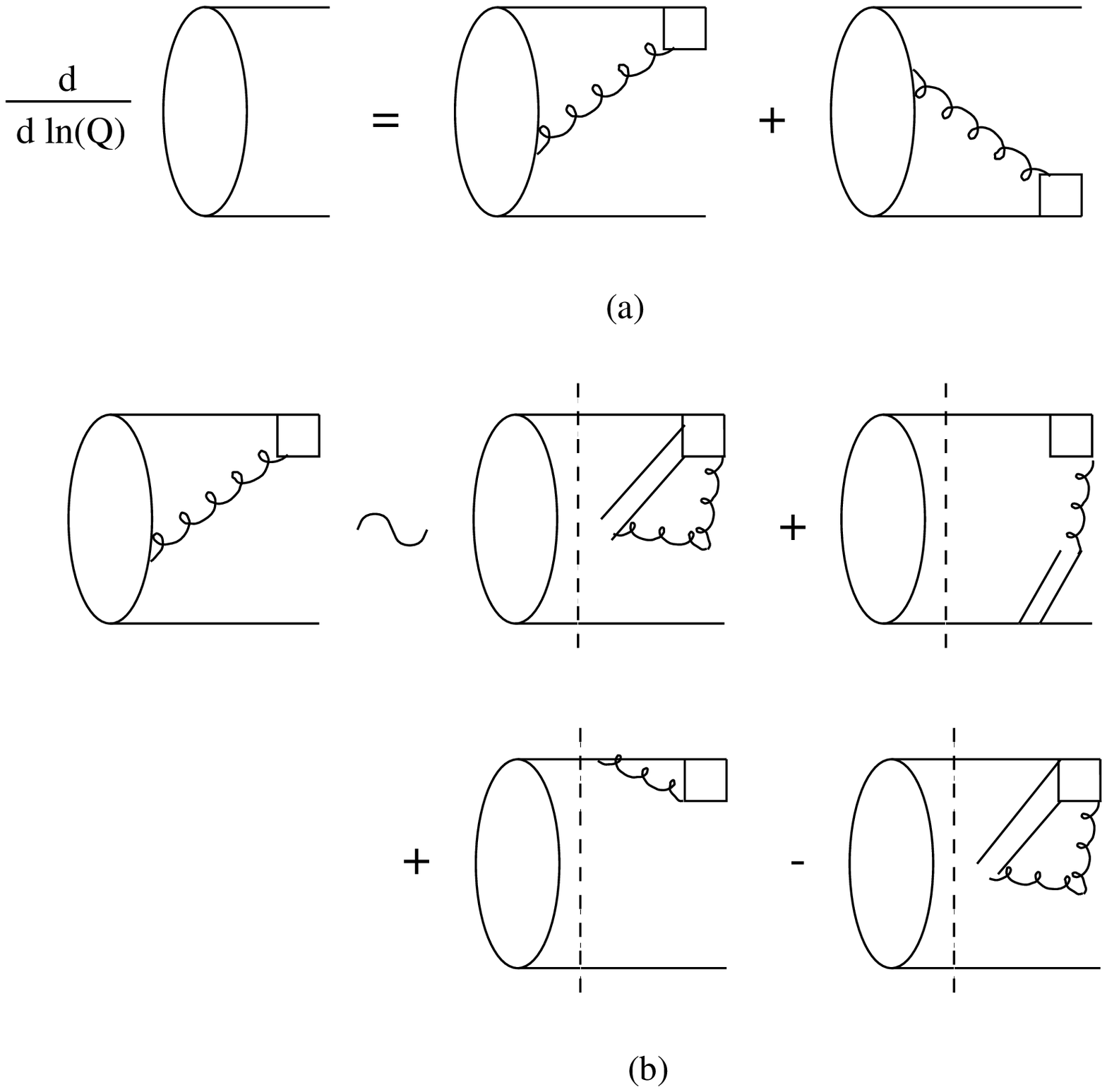}
\vspace{.2in}
\caption{(a) The derivative $d{\cal P}/d\ln Q$ in the axial gauge.
(b) The $O(\alpha_s)$ functions $K$ and $G$. }
\end{figure}

\begin{figure}[ht]
\includegraphics[angle=-90,width=5in]{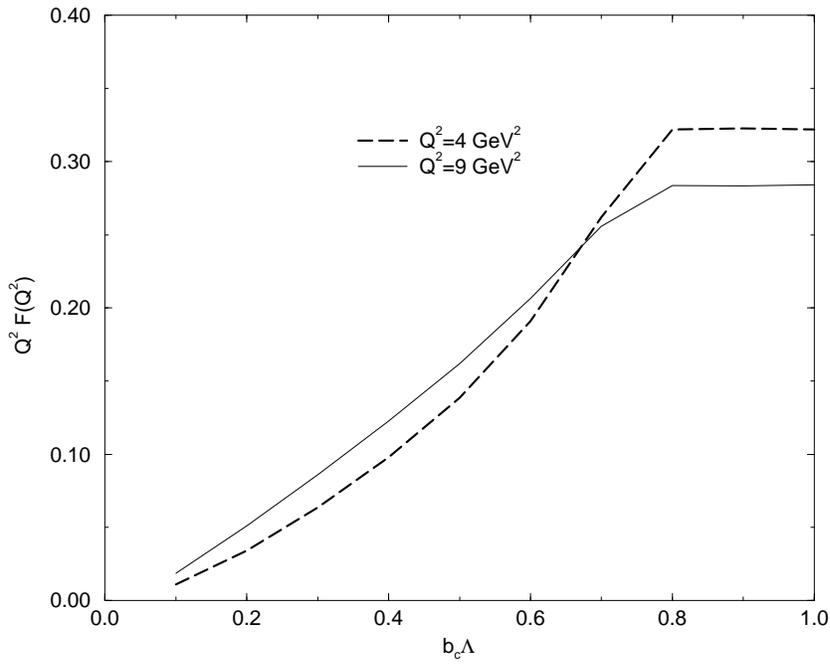}
\vspace{.2in}
\caption{Dependence of $Q^{2}F_{\pi}(Q^{2})$ on $b_{c}$ from
$\phi$ in Eq.~(\ref{phie}) for $Q^2=4$ GeV$^2$ (dashed
line) and for $Q^2=9$ GeV$^2$ (solid line).}
\end{figure}

\newpage
\begin{figure}[ht]
\includegraphics[angle=-90,width=5in]{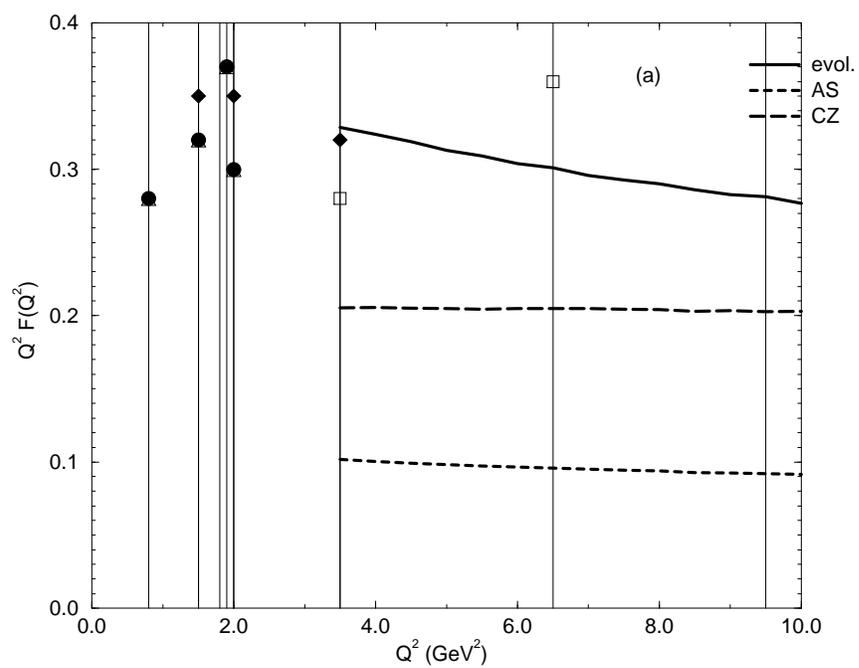}
\vspace{.2in}
\caption{Dependence of $Q^{2}F_{\pi}(Q^{2})$ on $Q^2$ from $\phi^{AS}$
(dotted line), form $\phi^{CZ}$ (dashed line), and from $\phi$ in
Eq.~(\ref{phie}) (solid line). }
\end{figure}

\newpage

\begin{figure}[ht]
\includegraphics[angle=-90,width=5in]{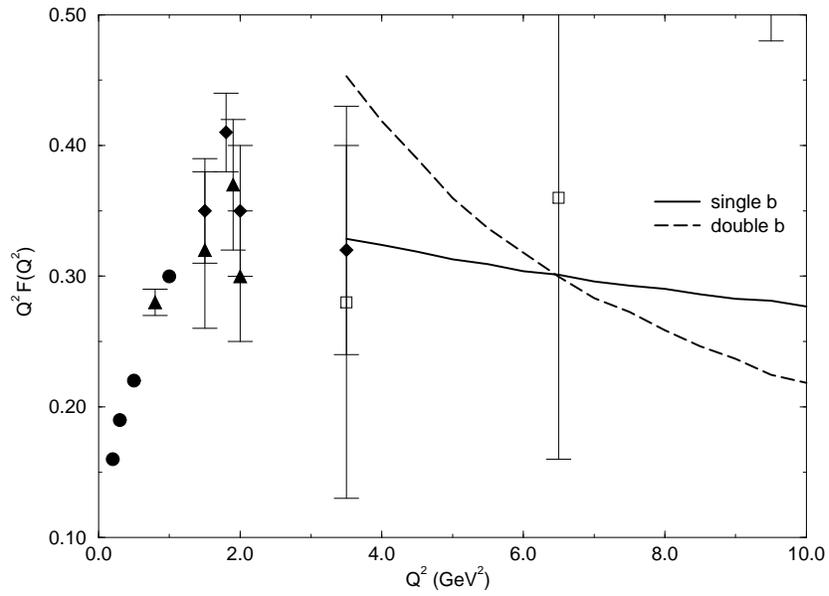}
\vspace{.2in}
\caption{Dependence of $Q^{2}F_{\pi}(Q^{2})$ on $Q^2$ for the use
of $\phi$ in Eq.~(\ref{phie}) derived from Eq.~(\ref{15})
(solid line) and from Eq.~(\ref{f5}) (dashed line). }
\end{figure}

\begin{figure}[ht]
\includegraphics[width=4in,height=6in]{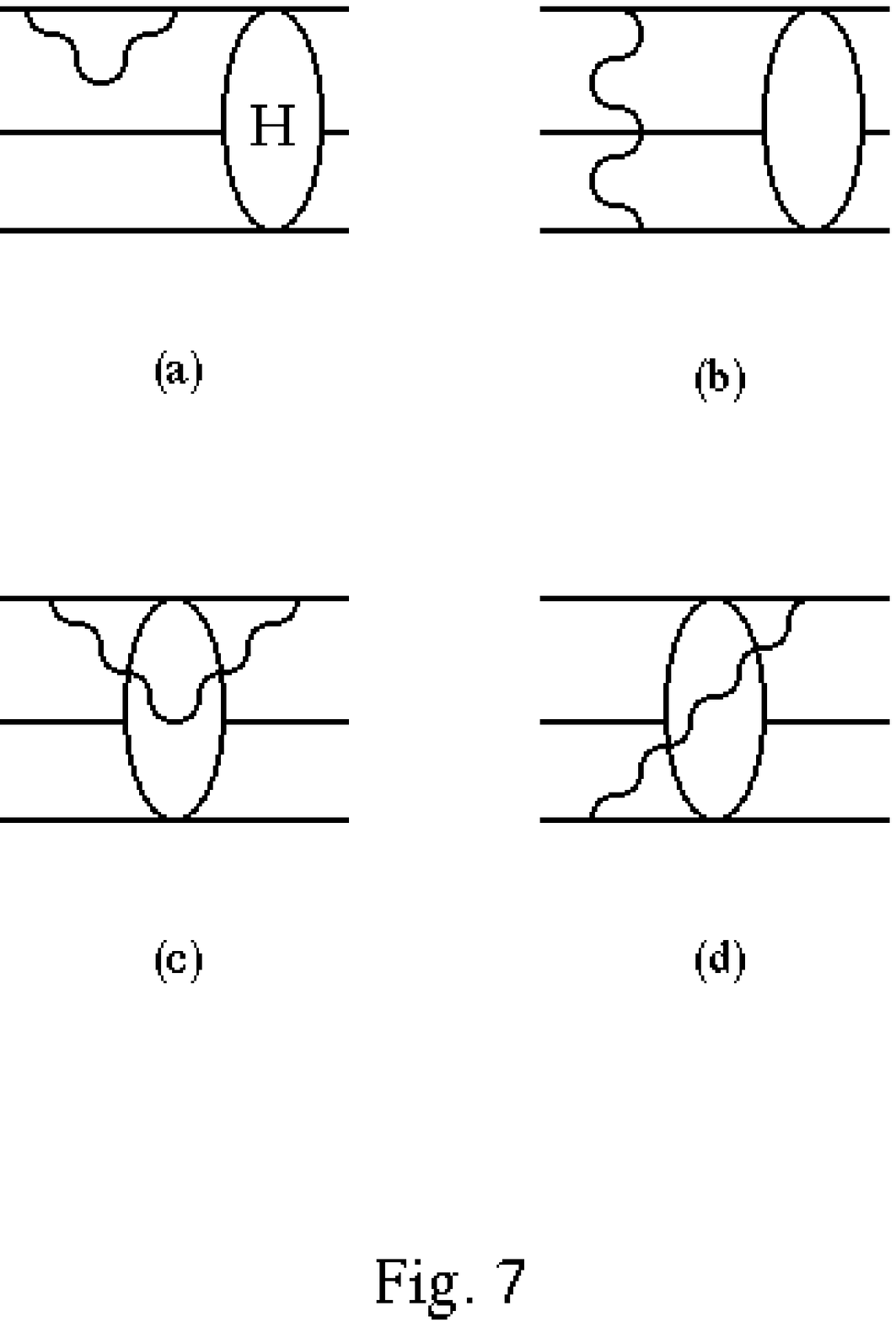}
\vspace{.2in}
\caption{(a)-(d) Radiative corrections to the basic diagrams for the
proton Dirac form factor.}
\end{figure}

\newpage
\begin{figure}[ht]
\includegraphics[width=4.5in,height=7in]{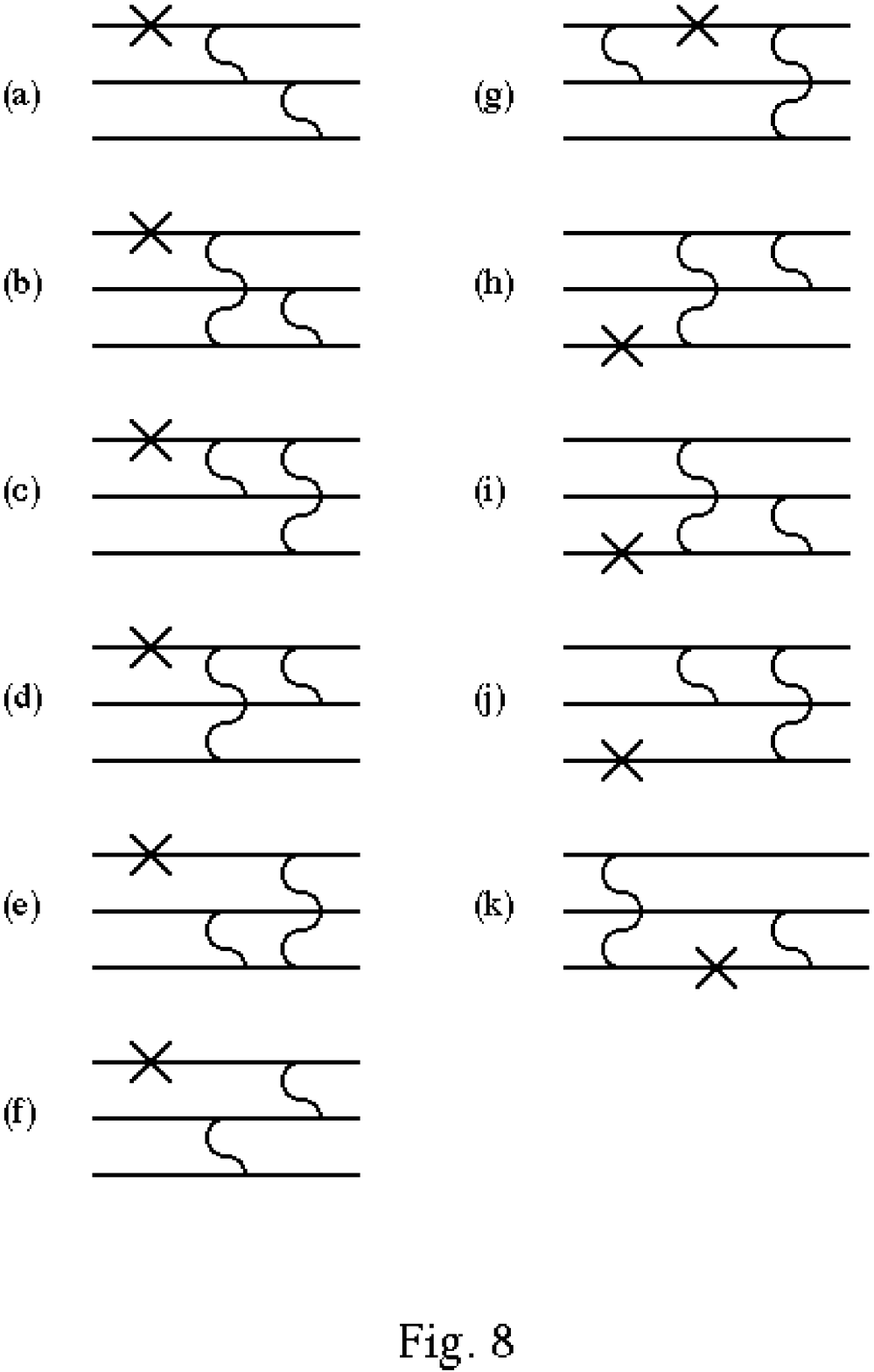}
\vspace{.2in}
\caption{Independent lowest-order hard scattering subdiagrams
for the proton Dirac form factor. The bottom line represents the $d$ quark.}
\end{figure}

\begin{figure}[ht]
\includegraphics[angle=-90,width=5in]{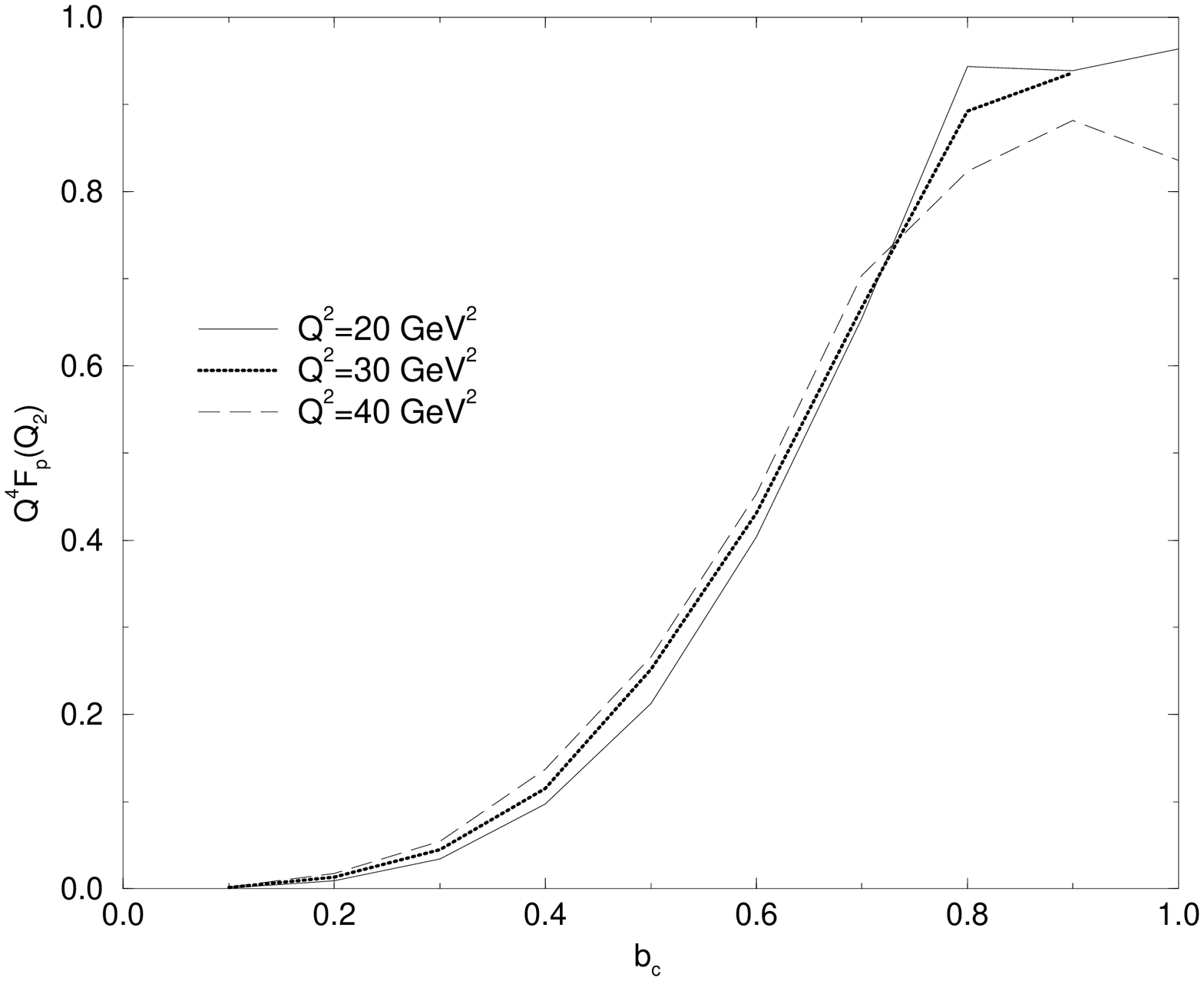}
\vspace{.2in}
\caption{Dependence of $Q^{4}F_{1}^{p}(Q^{2})$ on the cutoff $b_{c}$
from the KS wave function for $Q^2=20$ GeV$^2$ (dotted line), for $Q^2=30$
GeV$^2$ (dashed line), and for $Q^2=40$ GeV$^2$ (solid line). }
\includegraphics[width=6in,height=4in]{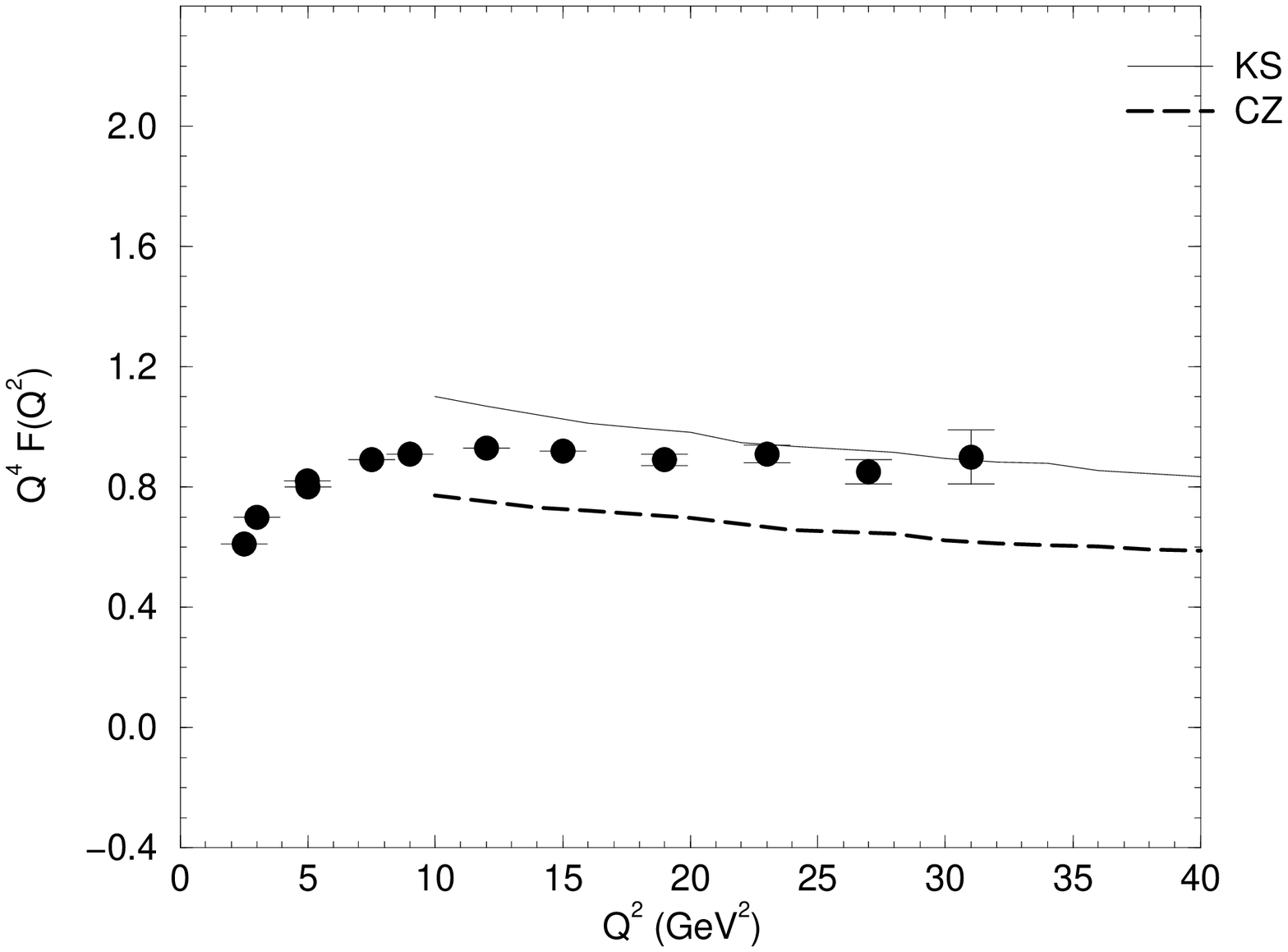}
\vspace{.2in}
\caption{Dependence of $Q^4F_1^p(Q^2)$ on $Q^2$ from the
KS wave function (solid line) and from the CZ wave function (dashed line).
The experimental data are also shown.}
\end{figure}

\newpage

\begin{figure}[ht]
\includegraphics[]{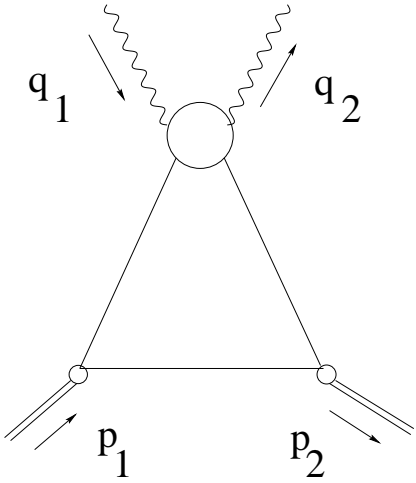}
\vspace{.2in}
\caption{The factorization picture for the 4-current correlator 
of pion Compton scattering.}
\end{figure}

\newpage

\begin{figure}[ht]
\includegraphics[angle=-90,width=6in]{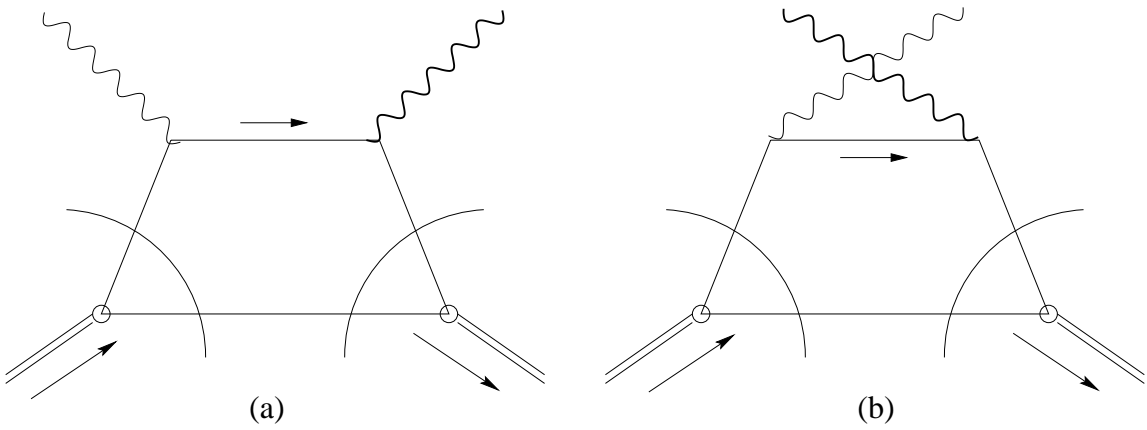}
\vspace{.2in}
\caption{(a) and (b) Lowest order perturbative contribution to the
spectral density for pion Compton scattering.
The dashed lines are on shell. }
\end{figure}
\newpage

\begin{figure}[ht]
\includegraphics[angle=-90,width=6in]{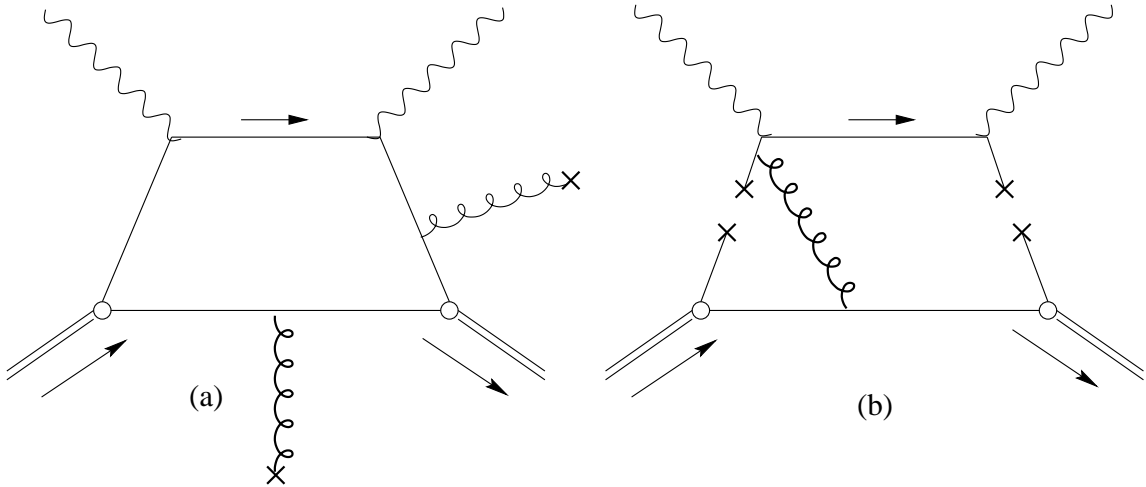}
\vspace{.2in}
\caption{(a) Typical gluon and (b) typical quark power corrections. }
\end{figure}

\newpage
\begin{figure}[ht]
\includegraphics[angle=-90]{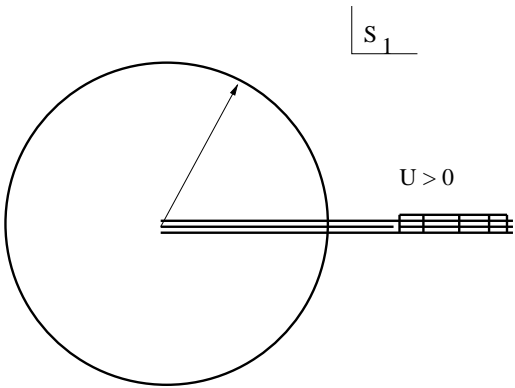}
\vspace{.2in}
\caption{The integration contours for the scalar amplitude in the $p_1^2$ plane, showing the $u=0$ threshold.}
\end{figure}

\begin{figure}[ht]
\includegraphics[angle=-90,width=7in]{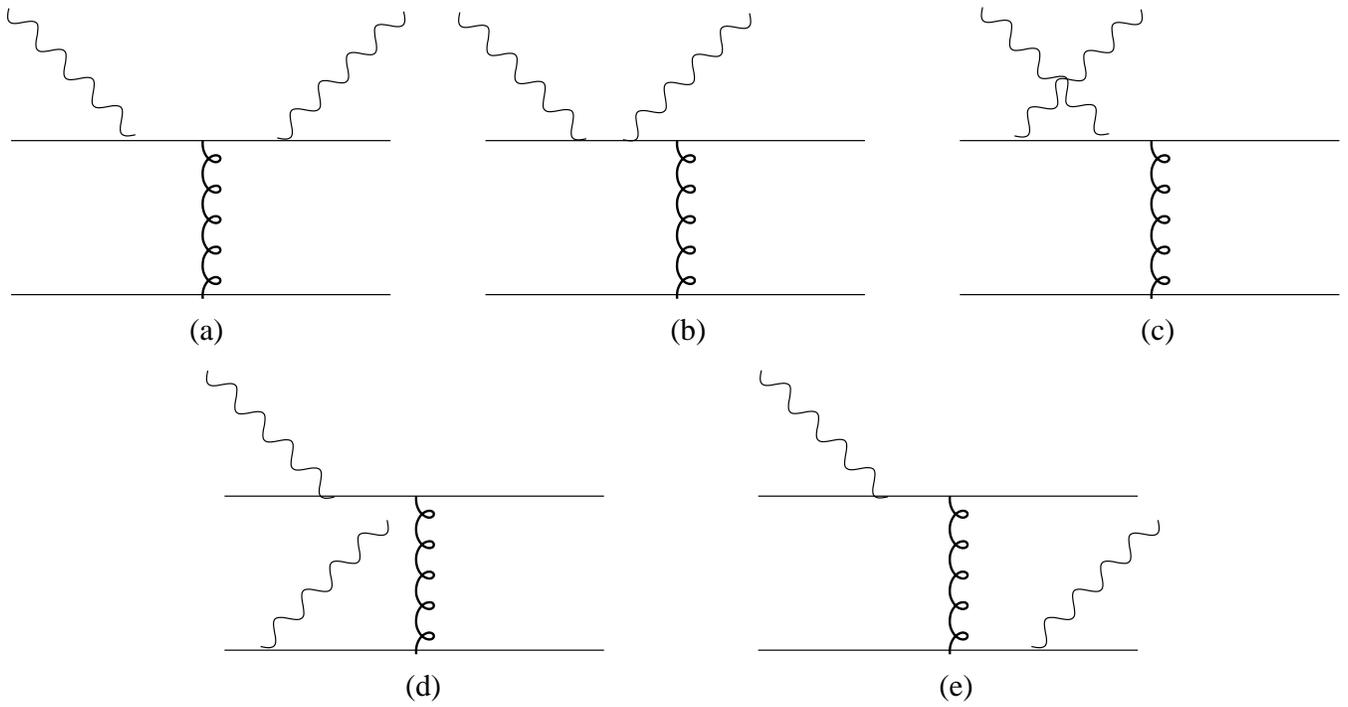}
\caption{Lowest-order diagrams for pion Compton scattering in
the PQCD approach.}
\end{figure}

\newpage

\begin{figure}[ht]
\includegraphics[angle=-90,width=6in]{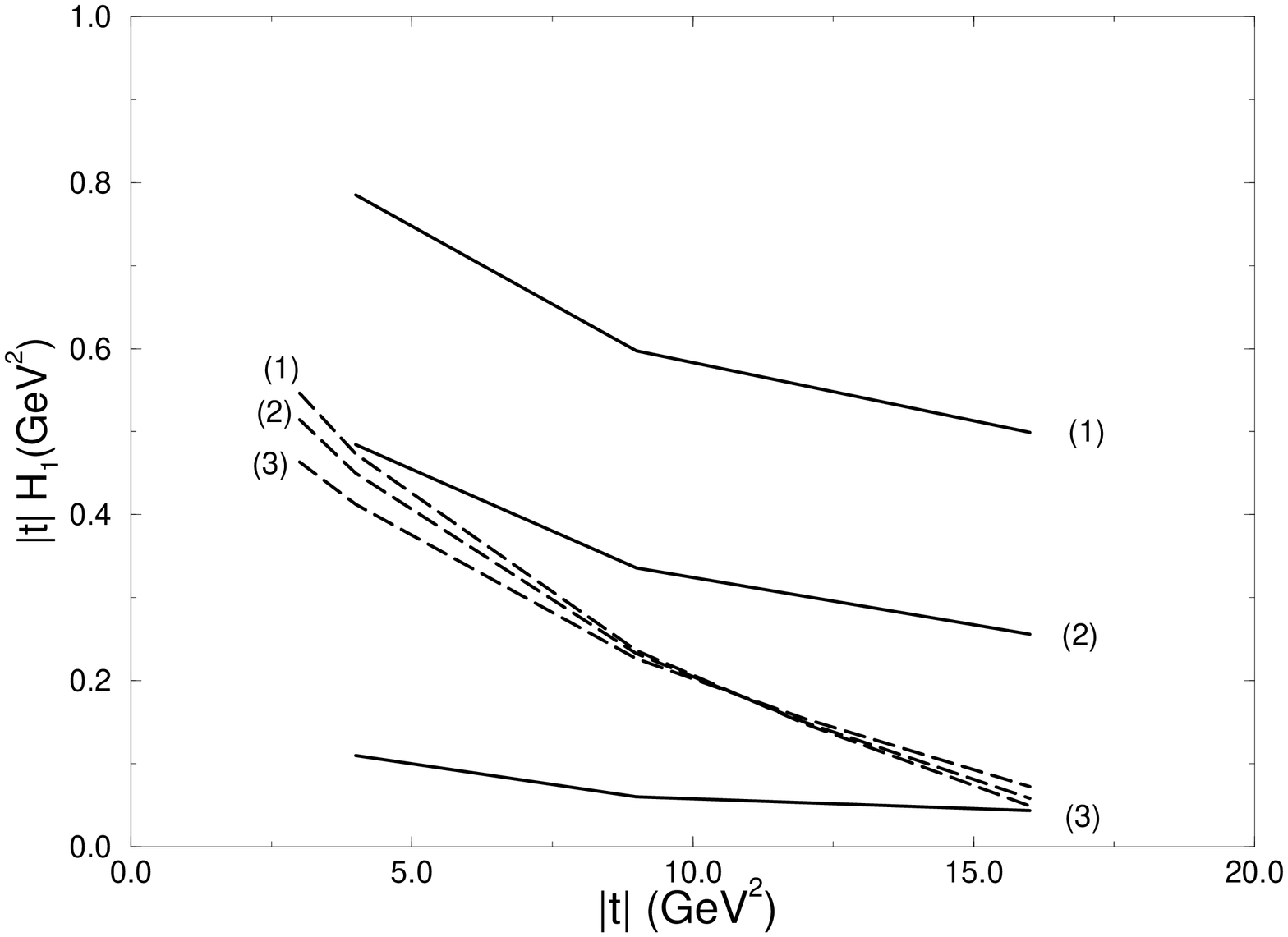}
\vspace{.2in}
\caption{Dependence of (a) $|t||H_1|$ on $|t|$
derived from the modified PQCD (solid lines) and from
QCD sum rules (dashed lines)
for (1) $-t/s=0.6$ ($\theta^*=50^o$) (2) $-t/s=0.5$ ($\theta^*=40^o$),
and (3) $-t/s=0.2$ ($\theta^*=15^o$).
}
\end{figure}

\newpage

\begin{figure}[ht]
\includegraphics[angle=-90,width=6in]{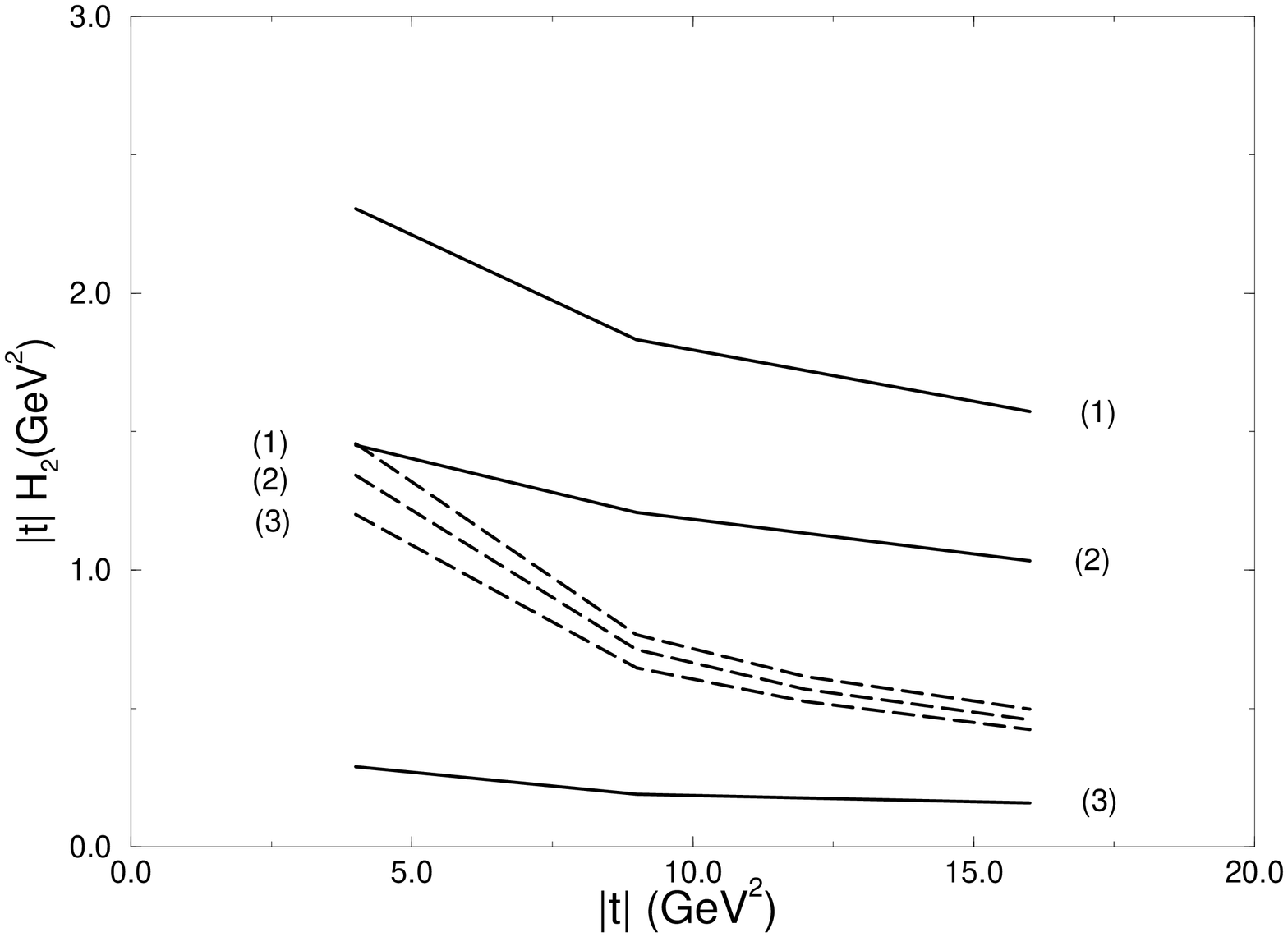}
\vspace{.2in}
\caption{$|t||H_2|$ on $|t|$
derived from the modified PQCD (solid lines) and from
QCD sum rules (dashed lines)
for (1) $-t/s=0.6$ ($\theta^*=50^o$) (2) $-t/s=0.5$ ($\theta^*=40^o$),
and (3) $-t/s=0.2$ ($\theta^*=15^o$).}
\end{figure}

\newpage
\begin{figure}[ht]
\includegraphics[angle=-90,width=6in]{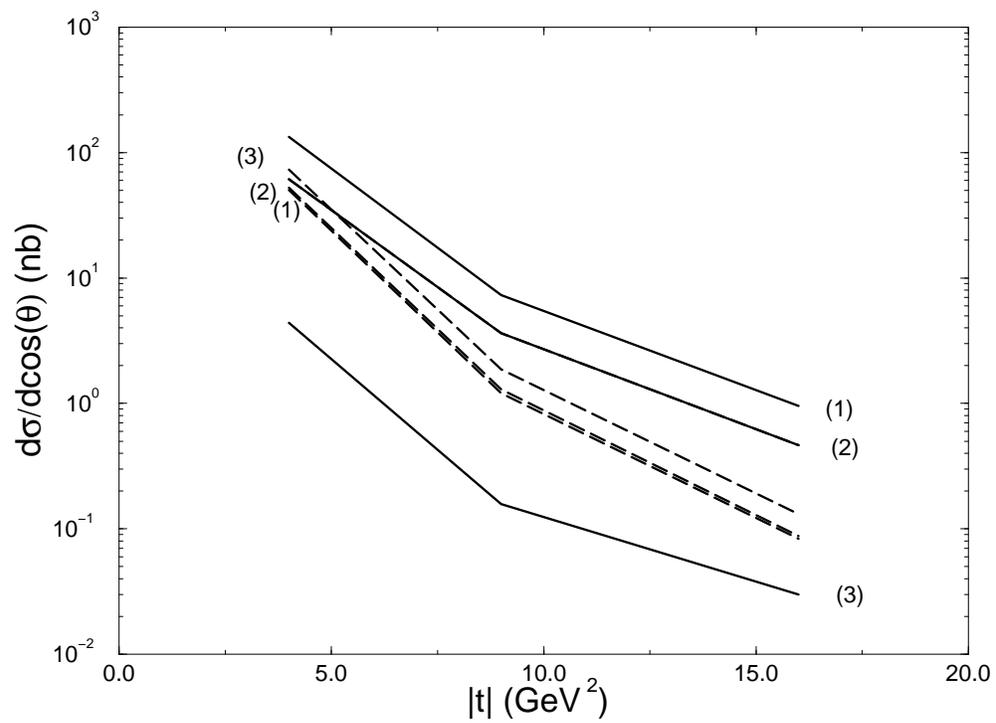}
\vspace{.2in}
\caption{ Dependence of $d\sigma/d\cos\theta^*$ on $|t|$
derived from the modified PQCD (solid lines) and from
QCD sum rules (dashed lines)
for (1) $-t/s=0.6$ (2) $-t/s=0.5$, and (3) $-t/s=0.2$. Note that
the curve (1) from sum rules is shown by a long-dashed line.}
\end{figure}

\newpage

\end{document}